\documentclass[12pt]{article}
\usepackage[T1]{fontenc}
\usepackage[latin9]{inputenc}
\usepackage{listings}
\usepackage{color}
\usepackage{array}
\usepackage{verbatim}
\usepackage{varioref}
\usepackage{units}
\usepackage{textcomp}
\usepackage{multirow}
\usepackage{amsthm}
\usepackage{amsmath}
\usepackage{amssymb}
\usepackage{graphicx}
\usepackage{esint}
\usepackage{grffile}
\usepackage[square,sort,comma,numbers]{natbib}

\makeatletter

\providecommand{\tabularnewline}{\\}

\numberwithin{equation}{section}
\numberwithin{figure}{section}
\theoremstyle{plain}

\definecolor{dkgreen}{rgb}{0, 0.6, 0}
\definecolor{mauve}{rgb}{0.58,0,0.82}
\definecolor{ltgray}{rgb}{0.9,0.9,0.9}
\def\xbar{\overline{x}}
\usepackage[font=footnotesize]{caption}
\usepackage{subfig}
\usepackage{courier}
\usepackage{adjustbox}
\usepackage{graphicx}

\@ifundefined{showcaptionsetup}{}{%
 \PassOptionsToPackage{caption=false}{subfig}}
\usepackage{subfig}
\makeatother

\usepackage[english]{babel}
\providecommand{\theoremname}{Theorem}

\newcommand*{\changes}[1]{#1}

\begin{document}


\title{Bayesian \changes{adaptive} bandit\changes{-based} designs using the Gittins index for multi-armed trials with normally distributed endpoints }

\author{Adam Smith$^{\rm a}$ and Sof\'{\i}a  S. Villar$^{\rm b}$
\thanks{Corresponding author. Email: sofia.villar@mrc-bsu.cam.ac.uk}\\
$^{a}${\em{Cambridge University, Cambridge}};\\
$^{b}${\em{MRC Biostatistics Unit}}\\
{\em University of Cambridge, School of Clinical Medicine}
}

\date{}
\maketitle

\begin{abstract}
\changes{Adaptive designs for multi-armed clinical trials have become increasingly popular recently in many areas of medical research because of their potential to shorten development times and to increase patient response. However,}
developing response-adaptive trial designs that offer patient benefit while ensuring
the resulting trial avoids bias and provides a statistically rigorous comparison
of the different treatments included is highly challenging.
In this paper, the theory of
\emph{Multi-Armed Bandit Problems} is used to define a family of near optimal \changes{adaptive} designs in the context of a clinical trial with a normally
distributed endpoint with known variance. Through simulation studies based on an ongoing trial as a motivation we report the operating characteristics (type I error, power, bias) and patient benefit of these approaches and compare them to traditional and existing alternative designs.
\changes{These results are then compared} to those recently published in the context of Bernoulli endpoints.
Many limitations and advantages are similar in both cases but there are also important differences, specially with respect to type I error control.
This paper proposes a simulation-based testing procedure to correct for the observed type I error inflation that bandit-based and adaptive rules can induce.
Results presented extend recent work by considering a \changes{normally distributed} endpoint,  \changes{a very common case in clinical practice yet mostly ignored in the response-adaptive theoretical literature}, and illustrate the potential advantages of using these methods in a rare disease context. \changes{We also recommend a suitable modified implementation of the bandit-based adaptive designs for the case of common diseases. }

\end{abstract}


\section{\label{sec:introduction}Introduction}

The medical and statistical communities
have long held as a `gold standard' for clinical trials the so-called
\emph{randomised controlled trial} (RCT), where patients are allocated to
a treatment arm with a fixed probability which is equal across all
arms and for all patients. This scheme ensures the trial is well-balanced,
eliminates possible sources of bias, and makes the results as sound
as possible. However, this design makes no concession to the wellbeing of
patients in the trial: in a $K$-arm RCT, on average $\tfrac{K-1}{K}$
of the patients will be assigned to a treatment other than the most effective one (if it exists).

This creates one of the foremost ethical concerns inherent in any clinical trial: the conflict between \emph{learning} \changes{(ensuring the selection of the best treatment)} and \emph{earning} \changes{(treating most patients effectively)}. The scientific
aim of a traditional RCT is to \emph{learn} about new treatments and identify
the most effective one. It is inevitable under this paradigm, however, that
a fixed number of patients will be given an inferior treatment. The research on adaptive methods for trial designs, such as response-adaptive randomisation methods, has developed as a response to this ethical dilemma, seeking to improve the \emph{earning} resulting from a trial while preserving its \emph{learning}.
The
challenge is to find  response-adaptive methods which improve patient
welfare during the trial, but do not allow extreme imbalance or bias to
hinder the statistical validity of the trial, and are conclusive enough
truly to influence future medical practice. 

The need to consider patients' wellbeing during the trial is particularly
acute in the case of a treatment for a rare disease. In this situation
the trial patients represent a high proportion of all those with the
disease, and a trial aiming solely to identify the most effective
treatment will benefit only the small number of patients remaining
to be treated after the end of the trial. The ethical concerns with
randomising patients onto an inferior treatment are most severe in
the case of a serious or life-threatening disease. \changes{Thus}, the motivation
for an adaptive trial design is arguably strongest in the case of
 life-threatening rare diseases \changes{such as the new types of rare cancers identified by the advances of genetics}. However, the challenges of maintaining statistical rigour are even
more acute when recruitable patients are sparse and sample sizes are
small.

The majority of the response-adaptive randomisation methods proposed in the literature use Bayesian learning \changes{ and a binary endpoint},
with information on the effectiveness of the treatments gained throughout
the trial deployed immediately, to increase the chances of patients in the trial receiving a better performing treatment (see e.g. \cite{trippa}). A limitation of these approaches is that they are myopic (they only make use of past information to alter treatment allocation probabilities) and hence they are not influenced at all by the number of patients that remain to be treated in the trial (nor by the expected  number of patients outside the trial).
An approach 
recently proposed and modified for addressing this limitation \changes{and developing ``\emph{forward looking algorithms}''} is to consider clinical trial design
within the framework of the \emph{Multi-Armed Bandit Problem (MABP)}.
The optimal solution to the classic MABP has been known since the
1970s %
{\cite{gittins-jones-1974}%
}, and those responsible for its solution saw clinical trials as the
``\emph{chief practical motivation}'' for their work %
{\cite[p. 561]{gittins-jones-1979}%
}; despite this, it has never been applied to a real life clinical trial. %
{\cite[pp. 2-3]{villar-a}%
}

In {\cite{villar-a} some of the benefits and limitations of applying the MABP
solution to clinical trials are explored, considering in particular the case where
the trial's primary endpoint is dichotomous (\emph{i.e.} the treatment
arms are modelled as reward processes by Bernoulli random variables).
The objective of the paper is two-fold. The first is to apply some of the considerations and techniques
of {\cite{villar-a} to \changes{define a response-adaptive Bayesian design} for a clinical trial whose primary endpoint
is normally distributed with known variance, \changes{ a case that has been less commonly studied in the response-adaptive literature}.  Specifically, we investigate whether
the same conclusions in terms of patient benefit and operating characteristics hold as in the case of trials with binary endpoints
and, since many trials do have normally distributed endpoints, in
this way we hope further to bridge the gap between MABP theory and
clinical trial practice. The second objective is to \changes{identify and address issues that may limit the use in pratice of the MABP-based designs considered in this paper. Specifically, we  }  consider in detail the level of bias and type I error rates observed under this setting and further suggest appropriate procedures to control them. Results are illustrated by simulations in the context of
a currently ongoing clinical trial: TAILoR trial, described in \citep{wason}.

The structure of this paper is as follows: In Section \ref{sec:mabp}
an overview of the general MABP with a continuous state variable and its solution \changes{for the special case of a normally distributed reward} is provided 
\changes{together with an adaptive patient allocation rule based on it}. Then, 
 Section \ref{sec:simulation-setups}
presents some simulations of two-armed and multi-armed trials
implementing bandit strategies for normally distributed endpoints and comparing them to alternative trial designs.
Section \ref{sec:conclusion} concludes with a discussion of our findings
and lines of further research.

\section{\label{sec:mabp}The classic Bayesian multi-armed bandit problem with a continuous state variable and
known variance}


Let $K\in\mathbb{N}$ and consider a collection $\{X_{k,t}:\, k=0,1,\ldots,K,\,\, t=0,1,2,\ldots\}$
of independent (real-valued) random variables, where for each fixed
$k$ the distributions of $X_{k,0},X_{k,1},X_{k,2},\ldots$ are identical
and parametrised by some unknown $\theta_{k}\in\mathbb{R}^{p}$. At
each time $t=0,1,2,\ldots,T-1$ we obtain a reward by choosing some distribution
$k\in\{0,\ldots,K\}$ and sampling from $X_{k,t}$. In the context
of a clinical trial, this corresponds to choosing the treatment allocation
of the $t^{\mathrm{th}}$ patient, and $X_{k,t}$ corresponds to the
endpoint observation for  patient $t$ on treatment $k$. 
In order to incorporate the \changes{adaptive learning} element into the model, we take
a Bayesian viewpoint and assume $\Theta_{k}$ is a random variable
taking the value $\theta_{k}$. We assign $\Theta_{k}$ a prior distribution
$\pi_{k}^{(0)}$, which is assumed to be a density function with respect
to Lebesgue measure. By Bayes' Theorem, the posterior density of $\Theta_{k}$,
having observed values $x_{k,i_{1}},\ldots x_{k,i_{n}}$ in $n$ independent
samples from $X_{k,i_{1}},\ldots,X_{k,i_{n}}$ after having treated $t$ patients, is
\[
\pi_{k}^{(t)}(\theta\left|x_{k,i_{1}},\ldots,x_{k,i_{n}}\right.)\propto\pi_{k}^{(0)}(\theta)\underset{j=1}{\overset{n}{\prod}}f_{k}(x_{k,i_{j}}|\theta),
\]
 where $f_{k}(\cdot|\theta)$ is the density of $X_{k,t}$ (with respect
to Lebesgue measure) {\cite
{kmt}
}. Note that we have used the subscript $i_j$ (for $j=1, \dots, n$) to emphasize that the sample of $n$ observations from distribution $k$ is a subset of the total number of sampling observations possible at time $t$.

Formally, the classic Bayesian MABP within this general setting is defined by formulating a Markov decision process as
follows. Consider each distribution (or arm) $k$ as a Markov process $B_{k}$ with
a Borel state space $(E_{k},\mathcal{E}_{k})$, by taking the state
\global\long\def\xikt{\xi_{k}(t)}
$\xikt\in E_{k}$ of $B_{k}$ at time $t$ to be the value $\tilde{x}$
of some chosen sufficient statistic $\tilde{X_{k}}$ for the posterior
density of $\Theta_{k}$, and updating the state every time we sample
from this arm. At each time $t=0,\ldots,T-1$ a decision variable $a_{k,t}\in\{0,1\}$
is chosen for process $B_{k}$ for each $0\le k\le K$, such that
exactly one arm receives action $1$ (is \emph{sampled}) and
all others receive action $0$ (their posterior density remains \emph{frozen}). If $a_{k,t}=0$
then $B_{k}$ is frozen (and so is its associated value for the sufficient statistics), thus  $\xi_{k}(t+1)=\xikt$ with probability
$1$. If $a_{k,t}=1$ then $\xi_{k}$ evolves according to a Markovian
transition kernel $\mathcal{P}_{k}$, \emph{i.e.} for any $A\in\mathcal{E}_{k}$
and $\tilde{x}_{0},\tilde{x}_{1},\ldots,\tilde{x}_{t-1},\tilde{x}\in E_{k}$
we have
\[
\begin{array}{rcl}
 \mathbb{P}[\xi_{k}(t+1)\in A|\xikt=\tilde{x},\,\xi_{k}(t-1)=\tilde{x}_{t-1},\ldots,\xi_{k}(0)=\tilde{x}_{0}]& = &\mathbb{P}[\xi_{k}(t+1)\in A|\xikt=\tilde{x}] \\
 & = & \mathcal{P}_{k}(\tilde{x},A).
\end{array}
\]
The transition kernel $\mathcal{P}_{k}$ is 
a density $p_{k}$
(with respect to Lebesgue measure on $\mathbb{R}$): 
\begin{equation}\label{transition}
 p_{k}(\tilde{x},\tilde{x}\star y)=\underset{\mathbb{R}^{p}}{\int}f_{k}(y|\theta)\pi_{k}^{(t)}(\theta|\tilde{x})\mathrm{d}\theta,
\end{equation}
where $\tilde{x}\star y$ denotes the updated value of $\tilde{X}_{k}$
if $y$ is the next value sampled.

If the process $B_{k}$ is sampled at time $t$ we earn the random reward
$R_{k}(\xikt,\xi_{k}(t+1))$, where $R_{k}:E_{k}^{2}\rightarrow\mathbb{R}$
is the \emph{reward function} of $B_{k}$. In the classic MABP this function is given by $R_{k}(\tilde{x},\tilde{x}\star y)=y$,
\emph{i.e.} the value of $R_{k}$ is the value taken by $X_{k,t}$.
We define $r_{k}:E_{k}\rightarrow\mathbb{R}$ as the \emph{expected}
\emph{reward} from the process in a given state %
{\cite
{puterman}%
}, given by
\begin{equation}\label{reward}
 \begin{array}{rcl}
r_{k}(\tilde{x}) & = & \mathbb{E}\left[R_{k}(\xikt,\xi_{k}(t+1))\left|\xikt=\tilde{x}\right.\right]\\
 &  & \,\\
 & = & \mathbb{E}\left[X_{k,t}\left|\tilde{X}_{k}=\tilde{x}\right.\right].
\end{array}
 \end{equation}

Let $E=E_{0}\times\cdots\times E_{K}$ be the joint state space
of the MABP, and $\xi(t)=(\xi_{0}(t),\ldots,\xi_{K}(t))\in E$ the joint
state vector of the MABP at time $t$. Let $\Pi$ be the set of
all \emph{feasible} sampling policies\emph{, i.e.} those in which
the decision at time $t$ depends only on past information and only sample one arm (or distribution) at a time.
Writing $a_{k,t}^{\pi}$ for the sequence of sampling decisions chosen by policy $\pi$,
the \changes{value function for the} classic MABP with a continuous state variable is 
\begin{equation}
V^*_d(\xi)=\underset{\pi\in\Pi}{\sup}\,\mathbb{E}\left[\left.\sum_{k=0}^{K}\sum_{t=0}^{T-1}d^{t}r_{k}(\xi_{k}(t))a^{\pi}_{k,t}\right|\xi(0)=\xi\right].\label{eq:state-value}
\end{equation}
Thus, the MABP is the problem of finding a policy
$\pi\in\Pi$ which maximises the value of the expected total discounted
reward of the sampling process. Notice that $d$ is a discount factor (i.e. $0 \le d < 1$) introduced for reasons of
tractability, so that the infinite horizon problem ($T = \infty$) can be considered.


One approach to \changes{solve} the MABP in \eqref{eq:state-value} would be via the dynamic programming equation
\begin{equation}
V^*_d(\xi)=\underset{i\in\{0,\ldots,K\}}{\max}\,\left\{ r_{i}(\xi_{i})+d\underset{E_{i}}{\int}\mathcal{P}_{i}(\xi_{i},\mathrm{d}y)R(\xi_{1},\ldots,\xi_{i-1},y,\xi_{i+1},\ldots,\xi_{K})\right\} .\label{eq:dp}
\end{equation}
Standard theory on Markov processes ensures that there is an optimal
solution to (\ref{eq:state-value}), and approximations to it may
be obtained using value iteration on (\ref{eq:dp}) %
{\cite
{ggw}%
}, but such an approach is computationally expensive, exploding
with the truncation horizon $T$ \changes{even for a small number of arms $K>3$}
{\cite
{villar-a}%
}.
For the infinite horizon MABP 
Gittins and Jones \cite{gittins-jones-1974} provided a theorem by which 
there
exists a function $\nu=\nu(B_{k},\tilde{x}_{k})$ such that at any time
the optimal strategy is to sample the process which has the highest
value of $\nu$.
%
There is a clear computational advantage to this approach: 
if we can compute a grid of values of $\nu$
for each bandit process, then the policy can be followed any number
of times 
by looking up values of $\nu$
for each process at each decision time.
Several proofs of the Index Theorem are given in \cite{ggw}. Gittins and Jones referred to $\nu$ as a \emph{dynamic allocation index}, \changes{but} this is now known widely
as the \emph{Gittins index}.

In the case where the MABP state space is discrete, as in the Bernoulli case,  values of $\nu$ can be looked up
from a matrix. With a continuous state space it is less clear that
 all the necessary calculations can be performed in advance. However,
 in most useful cases, including the normally distributed case,  the function $\nu(B_{k},\tilde{x}_{k})$ is
a linear function of \changes{$B_{k}$ and $\tilde{x}_{k}$ under} some discrete boundary
conditions; the discrete part can thus be calculated in advance as
a matrix of values \cite{ggw}.

For the finite horizon problem, which is the relevant case in the clinical trial context, \cite{villar-a} suggested an
index-based solution to the finite horizon MABP based on the {Whittle index} \cite{whittle}. However,
\changes{the {Whittle index} is omitted from the studies in this paper since}
in most trials its performance was near identical to that of the Gittins index (calibrated through the choice of the discount factor $d$) \changes{which further has a lower computational cost}.

\subsection{\label{sub:sub-normal-bandits}The Gittins index \changes{as a }Bayesian \changes{adaptive patient allocation rule for the normally distributed endpoint} 
 (with known variance)}

In this paper we \changes{consider clinical trials for which} 
the endpoint of each treatment arm $k$ is \changes{assumed to be }normally distributed
with unknown mean \global\long\def\mk{\mu_{k}}
$\mu_{k}$ and known variance \global\long\def\sk{\sigma_{k}^{2}}
$\sk$. We therefore consider
the rewards from arm $k$ to be an independent identically distributed
(\emph{iid}) sequence $X_{k,t}\overset{\mathrm{iid}}{\sim}N(\mk;\sk)$,
and $\mk$ is given a prior distribution $\pi_{k}^{(0)}$, which we
will take to be the improper uniform distribution on the whole real
line. \changes{The uniform prior distribution assumption will allow us to isolate the effect on
 patient welfare and other relevant statistical properties of the MAB adaptive design alone, i.e., without the use of prior (historical) data.}
 Let $f(\cdot|\mu;\sigma^{2})$ denote the density of a $N(\mu;\sigma^{2})$
distribution. If we have observed $n$ independent samples $x_{k,i_{1}},\ldots,x_{k,i_{n}}$
from $X_{k,i_{1}},\ldots,X_{k,i_{n}}$, then, writing \global\long\def\xknbar{\overline{x}_{k,n}}
$\xknbar=\tfrac{1}{n}\sum_{j=1}^{n}x_{k,i_{j}}$ for the sample mean,
by Bayes' Theorem the posterior density of $\mu_{k}$ at time $t$ is $\pi_{k}^{(t)}(\mu_{k}|x_{k,i_{1}},\ldots,x_{k,i_{n}})\sim N\left(\xknbar;\frac{\sk}{n}\right).$
A sufficient statistic for the posterior distribution
of $\mk$ is $(\xknbar,n)$, thus the state vector of process $B_{k}$ in this case
 will be $(\xbar,n)$ where $n$ is the number of observations so far
sampled from arm $k$ after having treated $t$ patients, and $\bar{x}$ is the mean of these observations.


As explained in \cite{ggw} for the MABP with normally distributed rewards with known variance, the indices $\nu(\xbar,n;\sk,d)$
can be written as follows %
\begin{equation}\label{indices}
\nu(\xbar,n;\sk,d)=\xbar+\sigma_{k}\nu(0,n;1,d).
  \end{equation}

Therefore, to implement the Gittins index policy at very low computational cost
it suffices to calculate in advance the values of $\nu(0,n;1,d)$.
This can be done to a good accuracy using value iteration on \eqref{eq:dp}
in the case of the two-armed bandit calibration setup. Details are
given in \cite[pp. 131-162]{amaral} and \cite[Chapters 7 \& 8]{ggw}. Computational results for this case were first computed in \cite{jones}.
The values of
the indices $\nu(0,n;1,d)$ used in this paper have been interpolated
from the tables printed in \cite[pp. 261-262]{ggw}.
\begin{figure}[b]
\includegraphics{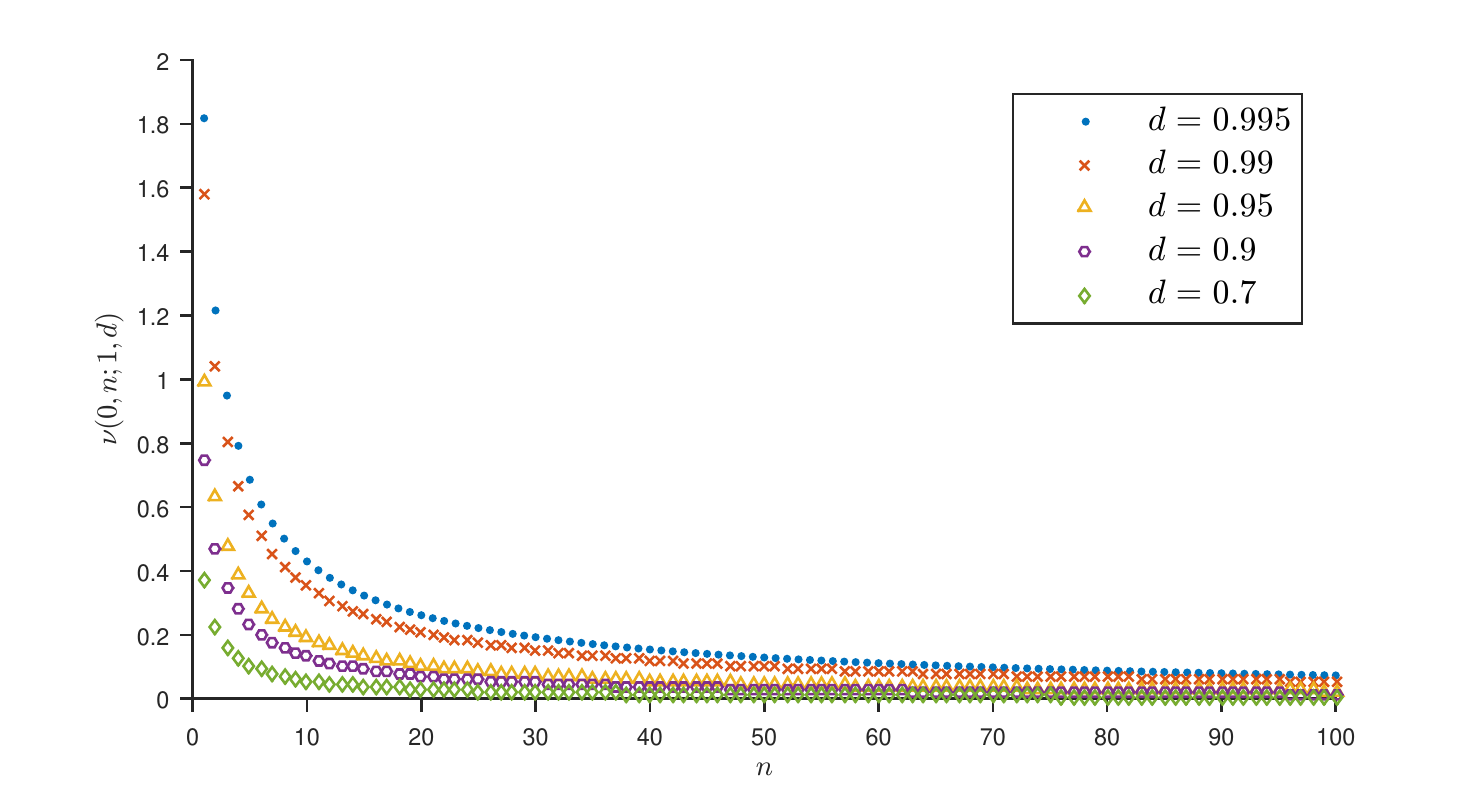}\caption{Gittins Index values (normal reward process, known variance) for various
discount factors $d$\label{fig:normal-gi-vals}}
\end{figure}

Figure \ref{fig:normal-gi-vals} shows the values of the indices $\nu(0,n;1,d)$
for a range of discount factors $d$. In \cite{gw}, the \emph{learning} component
of the index is defined as the difference between the index
value and the expected immediate reward, which
for this MABP corresponds to the reward
from sampling an arm with posterior mean $\xbar$ from previous
samples, i.e. simply $\xbar$. 
Therefore, $\sigma_{k} \nu(0,n;1,d)$ can be interpreted
as a measure of the \emph{learning} reward associated with continuing sampling
an arm which has already been sampled $n$ times. Figure \ref{fig:normal-gi-vals}
illustrates clearly that $\nu(0,n;1,d)$ increases with $d$, since
a larger discount factor puts greater value on future rewards and
increases the value of learning. However, for any choice of $d$,
the value of learning drops very quickly as $n$ increases; in the
limit as $n$ tends to infinity, the value of learning tends to $0$
and the sample mean converges by the Law of Large Numbers, so the
index tends to the true value of the parameter $\mu_{k}$.

\changes{The Gittins index solution for the case of both $\mu_{k}$ and  $\sigma_{k}$ being unknown exists and is  similar to that in \eqref{indices}. The difference is that the model requires a joint prior distribution on both parameters and the known variances in \eqref{indices} are replaced by sample variances.}

\subsection{\label{sub:sub-normal-bandits2} Some considerations specific to the use of bandit strategies in a clinical trial context }

In {\cite{villar-a} simulation results comparing a number of alternative patient allocation rules to index-based solutions for trial scenarios with dichotomous endpoints were provided.
The authors conclude that, alongside the clear advantages, there are
a number of limitations to the use of the Gittins index as an
allocation mechanism for clinical trials. Some of these disadvantages are still going to be an issue in the normally distributed case. The endpoint needs to be immediately observable so that index rules can be applied. This means that a patient in the trial cannot be treated
until all previous outcomes have been observed. \changes{This is a strong limitation that affects all adaptive designs in general and not only MAB-led designs. In practice,} this limits the speed
at which new patients can be recruited to the trial; however,
this may be less problematic in a rare disease context, where the
rate of patient recruitment is likely to be slow already. \changes{Applying the adaptive algorithms in batches of patients rather than patient after patient is a way of acknowledging and partially addressing this issue \cite{villar-b, perceht et al}.}

Another limitation still present is that
the allocation
of treatments in a Gittins index-based design is highly deterministic, which can lead
to the introduction of different sources of bias. As explained in \cite{atkinson-biswas}, randomisation
prevents the trial results from being influenced by ``secular trends
in the population's health and our ability to measure it, in the quality
of recruits to the trial and in the virulence of a disease.'' In
trials where the clinician can influence which patient receives the
next treatment, so-called \emph{selection bias} (the ability of the
experimenter to predict which treatment will be allocated next) can
influence the results.
These extrinsic bias effects are absent from the simulations
next presented and from those in \cite{villar-a}, but could have a significant impact when
deterministic rules are used on trials with real populations. Recent work \cite{villar-b} addresses this particular limitation proposing a simple modification of the Gittins index rule for the Bernoulli case that is randomised.
\changes{Notice that the lack of randomisation of the resulting patient allocations is a limitation shared with most bandit-based algorithms, even those that introduce random terms in their definitions as e.g., \cite{auer-et-al} or \cite{glazebrook}.}

For other limitations and also for the patient-benefit advantages of index-based designs reported in \cite{villar-a}, the magnitude or even their existence requires careful consideration.
This is the case for the possibility of introducing intrinsic sources of bias. 
Response-adaptive trials in general can result in biased estimate of a treatment's
outcomes. For example, in a two-arm trial scenario \cite{villar-a} found that the use of the Gittins index introduced a significant negative
bias in the estimate of treatment outcomes; the
magnitude of the bias is greatest for inferior treatments (since they
are more likely to be dropped early in the trial) and the treatment effects are likely to be overestimated.
Similar considerations apply with respect to the resulting rates of type I error (a false positive result, \emph{i.e. }incorrectly rejecting
the null hypothesis $H_{0}$) and  of type II error (failing
to detect that an experimental treatment is effective, \emph{i.e.
}incorrectly accepting $H_{0}$). \cite{villar-a} reported that the index-based designs achieved a level of statistical power that was far below the level of an RCT with the same number of patients $T$ and also that control of the type I error rate required adjusting the statistical test to correct for its conservativeness (i.e. moderate deflation).

\changes{An important contribution of this paper is to assess the extent to which further considerations different from the ones mentioned above apply to the normally distributed endpoint. In particular, assessing how important the bias and statistical error levels are in the normally distributed case, and suggesting how to control for the type I error rate at a desired level, are two of the main contributions of this work.}

\section{\label{sec:simulation-setups}Simulation studies}

In this section we evaluate the performance of a
range of patient allocation rules in a clinical trial context,
including  bandit-based solutions using the Gittins index.
As a case study for simulations we shall use a generalisation of the
currently ongoing TelmisArtan and InsuLin Resistance in HIV trial
(\emph{TAILoR trial}), which is described and also used as a case study in \cite{wason}. See also \cite{magirr} for discussion of the design
of the TAILoR trial.

The TAILoR trial is a one-sided test of $K$ experimental treatments against
a control treatment (\emph{i.e.} testing for superiority). Treatment
$k$ is assumed to have endpoint outcomes $X_{k,t}\overset{\mathrm{iid}}{\sim}N(\mk;\sigma^{2})$,
for $k=0,1,\ldots,K$ (where $k=0$ is the control treatment), and
$\sigma^{2}$ is known and common to all treatments. Setting $\delta_{k}=\mk-\mu_{0}$,
the global null hypothesis is $H_{0,G}:\delta_{1},\ldots,\delta_{K}\le0$
and the alternative hypotheses are $H_{1,k}:\delta_{k}>0,\,\,\,\,1\le k\le K$.

We focus on the following: statistical
power $(1-\beta)$; type-I error rate ($\alpha$); expected proportion
of patients in the trial assigned to the best
treatment ($\mathbb{E}p^{*}$); the \emph{Expected Outcome (}$\mathbb{E}$\emph{O)}
defined as the mean patient outcome across the trial realisations; and, for the two-arm case, bias in the maximum
likelihood estimate of treatment effect associated with
each decision rule.

For testing these hypotheses we shall use the following test statistics
\[
Z_{k}=\frac{\overline{X}_{k}-\overline{X}_{0}}{\sigma\sqrt{\tfrac{1}{n_{k}}+\tfrac{1}{n_{0}}}}
,\qquad k=1,\ldots,K,
\]
where $n_{k}$ is the number of sample observations taken from arm
$k$ and $\overline{X}_{k}$ is the sample mean of arm $k$. Under the assumption that the $n_{k}$'s are independent and identically distributed samples, these $k$ test statistics will follow a normal distribution with mean ${\delta_{k}}/{\left(\sigma\sqrt{\tfrac{1}{n_{k}}+\tfrac{1}{n_{0}}}\right)}$ and variance 1. In the case of a two-arm trial with one experimental treatment
to be tested against a control, this simplifies to the case of a standard
$z$-test using a univariate normal distribution. For the multi-armed case we will consider the joint distribution of $Z_{1},\ldots,Z_{k}$ and use a critical value ${C_{\alpha}} $ that controls the \emph{Family-Wise
type I Error Rate (FWER)}, defined as $\mathbb{P}\left[\{\mathrm{reject}\, H_{0,G}\}|H_{0,G}\right]$,
within a specified level $\alpha\in(0,1)$.

For each scenario we set the size of the trial $T$ to ensure that an RCT with equal randomisation achieves
a specified power $(1-\beta)$ to detect a specified effective treatment
difference $\delta^{(1)}$ between each arm and the control, while
controlling the FWER within $\alpha$. Because we are interested in the marginal type II error rate in a single
test, rather than a family-wise error rate,
we consider the marginal distributions rather than a joint
distribution to determine a required sample size per arm for an RCT. Following this rationale it can be computed that
the total required size of the RCT trial (i.e. across all arms) is
\begin{equation}
T=\underset{k=0}{\overset{K}{\sum}}n_{k}=(K+1)\left(\frac{2 \sigma^{2}(C_{\alpha}+z_{\beta})^{2}}{\left(\delta^{(1)}\right)^{2}}\right),\label{eq:trial-size}
\end{equation}
where $z_{\beta}$ is the $100(1-\beta)^{\mathrm{th}}$-percentile
of a standard $N(0,1)$ distribution. \changes{See appendix I for details of how ${C_{\alpha}} $  is determined and e.g., \cite{zhong} for a review of sample size calculation in RCTs.}

Following \cite{wason}, we
shall assume the variance in the outcomes is $\sigma^{2}=1$, and
specify the treatment difference to be detected as $\delta^{(1)}=0.545$
(chosen such that the probability of a patient given a treatment $k$
with $\delta_{k}=\delta^{(1)}$ having a better outcome than a patient
on the control treatment is $0.65$). We will consider the usual error rates of $\alpha=0.05$ and \changes{$\beta= 0.10$}.


In every scenario we consider the
following patient allocation procedures:

\begin{itemize}
\item \emph{Fixed Randomised (FR)}: for each patient, treatments are allocated
randomly with fixed probability $\tfrac{1}{K+1}$ across all treatments;
\item \emph{Thompson Sampling (TS)\cite{thompson1933likelihood}:} for each patient, treatments are allocated
randomly, where the probability $\pi_{k,t}$ of allocating treatment
$k$ to patient $t$ is proportional to the posterior probability
that treatment $k$ is the best, \emph{i.e.}
\[
\pi_{k,t}=\dfrac{\mathbb{P}\left[\underset{i}{\mathrm{max}}\,\mu_{i}=\mu_{k}\,|\, X_{k,t}\right]^{c}}{\sum_{l=0}^{K}\mathbb{P}\left[\underset{i}{\mathrm{max}}\,\mu_{i}=\mu_{l}\,|\, X_{l,t}\right]^{c}}
\]
where $c$ is a tuning parameter defined as $\tfrac{t}{2T}$ \changes{introduced to stabilise the resulting allocation probabilities \cite{thall-wathen}}. The
probabilities in the fraction are estimated by simulation at each
$t$
;
\item \changes{\emph{Upper Confidence Bound (UCB)}}: for the first $K+1$ patients,
patient $t$ is allocated treatment $k=t-1$; each patient $t>K+1$
is allocated the treatment $k$ with the highest value of the index
$\overline{x}_{k}+\sigma\sqrt{\frac{2\mathrm{ln}t}{n_{k}}}$, \changes{as proposed in
\cite{katehakis-robbins} and \cite{auer-et-al}};
\item \changes{\emph{Kullback-Leibler Upper Confidence Bound (KLU)}: for the first $K+1$ patients,
patient $t$ is allocated treatment $k=t-1$; each patient $t>K+1$
is allocated the treatment $k$ with the highest value of the index
$\overline{x}_{k}+\sigma\sqrt{\frac{2(\mathrm{ln}t+3(\mathrm{ln}\ \mathrm{ln}t))}{n_{k}}}$. This variant of UCB was shown in
\cite{cappe-et-al} to have improved asymptotic regret bounds compared to UCB};
\item \emph{Current Belief (CB)}: the next patient is allocated the treatment
with the highest posterior mean $\xbar_{k}$;
\item \emph{Gittins Index (GI)}: the next patient is allocated the treatment
with the highest value of the Gittins Index $\nu(\xbar_{k},n_{k};\sigma^{2},d)$,
where $d$ is the value of the discount factor;
\item \emph{Randomised Gittins Index (RGI)}: as first suggested in \cite{glazebrook}, the next patient is allocated
the treatment with the highest value of the semi-randomised index
$\nu(\xbar_{k},n_{k};\sigma^{2},d)+\tfrac{K+1}{n_{k}}Y_{t}$, where
$Y_{t}$ is a random variable sampled from the exponential distribution
with mean $\tfrac{1}{K+1}$ (this choice of randomisation element
is the same as that used by \cite{villar-a});
\item \emph{Randomised Belief Index (RBI)}: as first suggested in \cite{bather}, the next patient is allocated
the treatment with the highest value of the semi-randomised index
$\xbar_{k}+\tfrac{K+1}{n_{k}}Y_{t}$, where $Y_{t}$ is a random variable
sampled from the exponential distribution with mean $\tfrac{1}{K+1}$.\\

\noindent
\changes{For the multi-armed scenarios we have additionally considered the following rules: }
\item \emph{Trippa et al. Procedure (TP)}: for each patient, treatments
are allocated randomly, where the probability $\pi_{k,t}$ of allocating
treatment $k$ to patient $t$ is defined by
\[
\pi_{k,t}=\dfrac{\overline{\pi}_{k,t}}{\sum_{l=0}^{K}\overline{\pi}_{l,t}},
\]
where
\[
\overline{\pi}_{k,t}=\begin{cases}
\dfrac{\mathbb{P}\left[\mu_{k}>\mu_{0}\,|\,\overline{x}_{k,t-1}\right]^{\gamma_{t}}}{\sum_{l\ge1}\mathbb{P}\left[\mu_{l}>\mu_{0}\,|\,\overline{x}_{l,t-1}\right]^{\gamma_{t}}}, & \, k\ge1\\
\frac{1}{K}\mathrm{exp}\left[\underset{k=1,\ldots,K}{\mathrm{max}}\left[n_{k,t}-n_{0,t}\right]^{\eta_{t}}\right], & \, k=0,
\end{cases}
\]
in these simulations we have considered $\gamma_{t}=3\left(\tfrac{t}{T}\right)^{1.75}$ and $\eta_{t}=0.25\left(\tfrac{t}{T}\right)$,
as in \cite{trippa}. \changes{Note that this procedure is only considered for multi-arm trials because by design its allocation to the control arm will closely follow that of the best experimental arm. }%

\item \emph{Controlled Gittins (CG)}: 
each patient is randomly assigned the control treatment with (fixed)
probability $\tfrac{1}{K}$. If the patient is not randomly assigned
to the control group in this way, then she is assigned to the treatment with the greatest Gittins index. Although CG deviates from
the optimality of GI, it was still found in \cite{villar-a} to offer a significant improvement
in patient welfare over FR; moreover, it largely combatted the issue
of reduced power. In fact, when there existed a clear superior treatment among the $K$ arms it was found
to achieve even higher power than FR.
\item \changes{\emph{Controlled UCB (CUC)}: a variant of UCB with the control allocation protected as in CG above.}\\

\noindent
\changes{In all scenarios we also include ``batched'' versions of the Bayesian rules in which the allocation probabilities are updated after a block of $b$ patients are treated instead of after every patient. This idea was implemented in \citep{villar-b,perceht et al} as a means of overcoming the practical limitations imposed by the assumption of immediate outcome observability of these algorithms. Their inclusion is intended to more closely replicate the constraints of a real life clinical trial without fully sequential design. Specifically, we consider:}

\changes{\item \emph{Batched Thompson Sampling (TSB)} and \emph{Batched Trippa et al. Procedure (TPB)}: as per TS and TP above, but the allocation probabilities $\pi_{k,t}$ are only updated after every 20 patients. The size of the batch ($b=20$) was chosen to illustrate the effects of a moderate delay in relation to the first two trial sizes considered in this paper and a more severe delay for the trial size reflecting a rare disease scenario. Note that none of the trial sizes assumed are exactly divided by $b=20$. This leaves a few remaining patients which are allocated using the allocation probabilities resulting after observing the outcomes of last block. }
\end{itemize}

\changes{In every scenario considered and for every procedure we assumed that the prior for the $\mu_k$ parameters is the improper uniform distribution on the whole real line. Notice that a fully Bayesian approach to the design could make use of historical data existing before the trial through appropriate choice of the prior distribution. In this paper we have chosen to use an uninformative prior to make results comparable to the case study in \cite{wason} and to isolate the effects of the adaptive designs in the different performance measures.}

For the rules that are based on the Gittins index values there is an obvious ethical concern around the choice of a discounting factor
when calculating the indices: clearly current and future patients'
wellbeing should be valued equally. So for scenarios with large
sample sizes we will take $d$ close to $1$, usually $d=0.995$.
In the case of a rare disease, if it is known that not more than $N$
patients will ever be treated, the value
of $d$ could be chosen so that $d^{N}\approx0$, to ensure that the possibility of treating patients beyond $N$ has little
impact on current decisions. \changes{By doing this the current estimation of the patient population could be used to indirectly affect the choice of trial design. }
In all trial designs and in all simulations,
ties among index values are broken randomly.


\subsection{\label{sub:two-arm}Two-arm trial}

We first simulate the TAILoR trial with one experimental arm to be
compared with a control treatment, \emph{i.e. }$K=1$. The trial is
implemented under $H_{0}$ with $\delta_1=\mu_{0}=\mu_{1}=0$ and under $H_{1}$
with $\mu_{0}=0,\,\mu_{1}=0.545$ (i.e. $\delta_1=0.545$). In both scenarios the common variance
is $\sigma^{2}=1$ and $d=0.995$.
The sample size considered is of $T=116$ patients. This size ensures a 5\% type I error rate (using $C_{\alpha}=1.645$ as a critical value) and$(1-\beta)=90\%$ power to detect a difference
of $\delta^{(1)}=0.545$ through a FR design.

\subsubsection{\label{sub:2arm-MC} Type I error control for adaptive designs}

For the adaptive allocation
mechanisms, including the Gittins index-based, the use of a critical value of $C_{\alpha}=1.645$ is found in simulations to
generate a type I error rate inflated above $5\%$. This is in stark difference to the type I error deflation reported in \cite{villar-a,villar-b}.
We will next explain this phenomenon in detail in terms of the GI rule but a similar logic applies for other adaptive rules.

In
each realisation of the trial, if one arm performs badly early on
and is dropped (or allocated with a very low probability), then the sample mean from this arm will not have a chance
to regress upwards to the mean (or do so more slowly), being therefore negatively biased.
Figure \ref{fig:2arm-bias-once} illustrates this in a typical GI
trial run under $H_{0}$, displaying the posterior mean $\xbar_{k,t}$
of the outcomes for each treatment arm $k$ after $t$ patients have
been allocated that arm. In this example, the control arm $k=0$ performs
badly early on in the trial, so is dropped with just $n_{0}=19$ patients,
leaving the trial's estimate of this treatment's outcomes negatively
biased. The experimental arm performs better early on, so is continued
and regresses to its mean; thus the trial's final estimate of this
treatment's effect is close to the true value of 0. The result is
that the test statistic takes the value $1.81>1.645$, so a hypothesis
test using the normal cut-off value of $1.645$ would generate a type
I error, incorrectly concluding the superiority of the experimental
arm.

\begin{figure}[h!]
\begin{centering}
\includegraphics{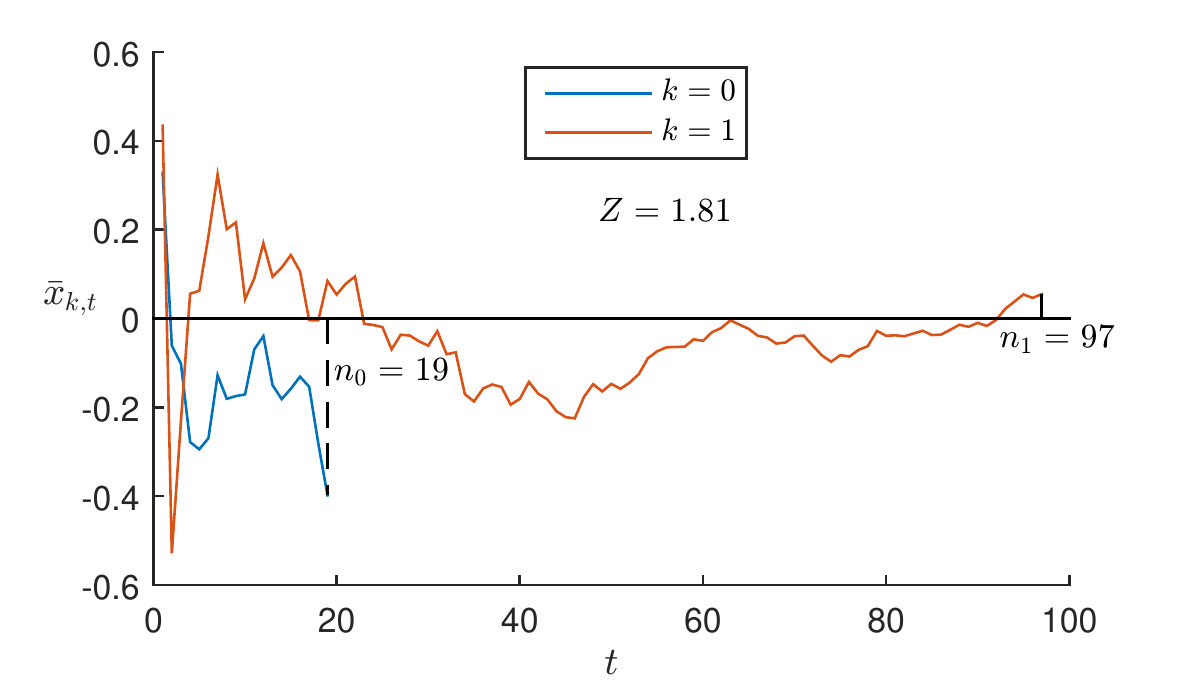}
\par\end{centering}

\caption{\label{fig:2arm-bias-once}The posterior mean $\bar{x}_{k,t}$ of
each treatment arm's outcomes after each patient in a typical GI trial
under $H_{0}$}

\end{figure}

In order to choose a more suitable critical value for the hypothesis
tests when using adaptive designs, we estimate the distribution of the test statistic $Z$ under each trial design
by a Monte Carlo simulation with $10^{4}$ repeats of the trial under
$H_{0}$. Figure \ref{fig:two-arm-hist} shows the observed empirical
distributions of $Z$ for the GI trials, implemented under
$H_{0}$ and under $H_{1}$. In each case, as well as a histogram
of the observed empirical distribution, also displayed is a curve
of the standard normal distribution which the test statistic is expected
to follow in a FR trial, for comparison. 

In Figure \ref{fig:two-arm-hist-gi-h0} we see that, in the GI trial
under $H_{0}$, the distribution of the test statistic is starkly
different from a normal distribution. The sample standard deviation
of $1.37$ is much greater than the standard deviation of $1.00$
in the FR case, and the heavier tails than a normal distribution correspond
to an inflated type I error rate when hypothesis testing is carried
out with the normal critical value of $1.645$.
Notice that because both left and right tails are heavier than the normal tails,
the inflation of type I error rate when testing hypotheses at the
normal cut-off value would be even greater in a two-tailed test.

The empirical cumulative distribution function evaluated at $1.645$
is $\hat{F}_{GI,H_{0}}(1.645)=0.89$, indicating that we might expect
a type I error rate of $11\%$ if hypothesis testing was carried out
with this critical value. Instead, the empirical $95^{\mathrm{th}}$-percentile
of the distribution is $C_{0.05}=\hat{F}_{GI,H_{0}}^{-1}(0.95)=1.951$,
marked on the histogram by a vertical dotted line. We will therefore
use this as the critical value for hypothesis testing in the GI trials
to control the type I error rate within $5\%$. 

Notice that the two peaks in the
frequency density arise from the two situations in which the estimate
of one arm's outcomes is negatively biased, and the other is unbiased:
the right-hand peak corresponds to an incorrect conclusion that the
experimental treatment is superior to the control treatment (the situation
illustrated in Figure \ref{fig:2arm-bias-once}), and the left-hand
peak corresponds to an incorrect conclusion of the opposite.

\begin{figure}[h!]

\subfloat[\label{fig:two-arm-hist-gi-h0}GI trial under $H_{0}$]{\includegraphics{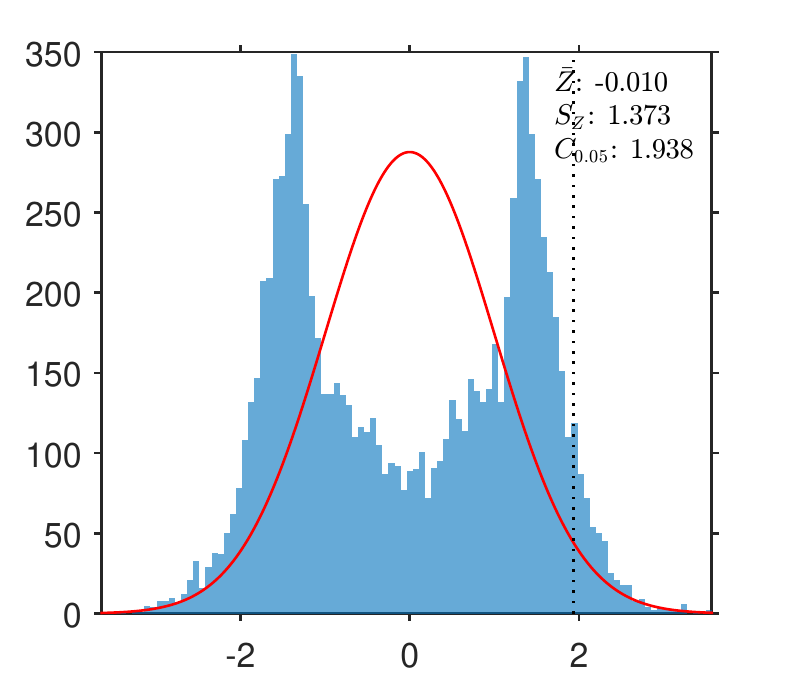}}\hfill{}
\subfloat[\label{fig:two-arm-hist-gi-h1}GI trial under $H_{1}$]{\includegraphics{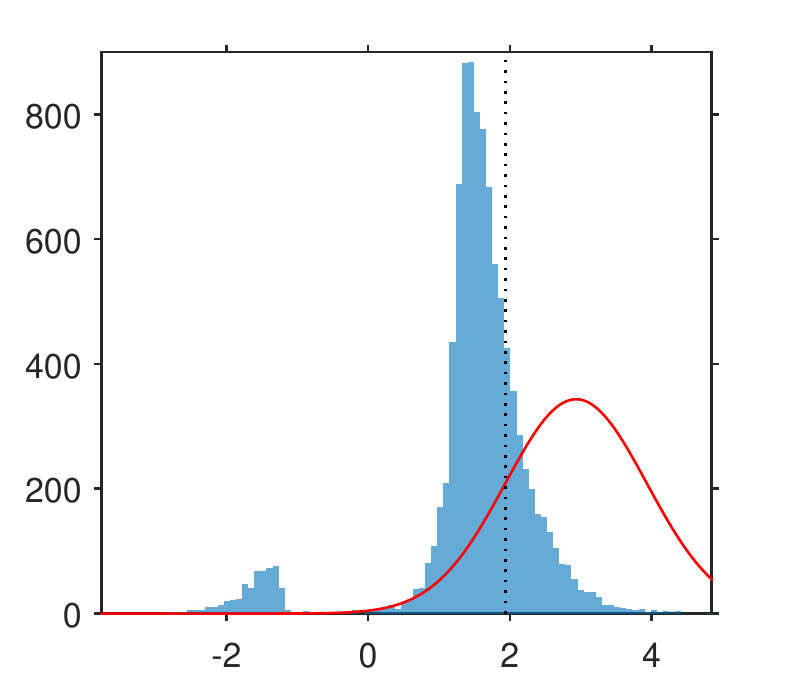}}
\caption{\label{fig:two-arm-hist}Histograms of empirical distributions of
the test statistic $Z$ in GI trials, implemented under each
hypothesis. Also marked is the standard normal distribution which
$Z$ should follow in the FR trial (red). The sample
mean $\bar{Z}$, standard deviation $S_{Z}$ and an empirical $95^{\mathrm{th}}$-percentile
$C_{0.05}$ have been calculated under $H_{0}$. The empirical $95^{\mathrm{th}}$-percentile
under $H_{0}$ will correspond to the critical value for hypothesis
testing, and is marked by a vertical dotted line on the histograms.}
\end{figure}

 Figure \ref{fig:two-arm-hist-gi-h1} illustrates that if the GI trial
is implemented under $H_{1}$ the bimodality of the distribution
of the test statistic is greatly reduced, but still present to some
extent. $\hat{F}_{GI,H_{1}}(1.951)=0.77$, \emph{i.e.} $77\%$ of
the distribution still lies to the left of the empirical critical
value of $1.951$, marked by a vertical dotted line; thus we expect
to observe greatly reduced power of around $23\%$ in the GI trials.
The (small) left-hand peak has a weight of $\hat{F}_{GI,H_{1}}(-0.5)=5\%$,
indicating that in $5\%$ of trials the superior arm is dropped
early on due to poor initial performance, and the trial has ended
up favouring the wrong arm.

Following the same procedure, $95^{\mathrm{th}}$-percentiles of the test statistic distribution
are estimated  for the other adaptive trial designs. Histograms
for the distributions of the test statistics in \changes{the other trial designs} are displayed in Appendix II
in Figure \ref{fig:app-2arm-hist}.
Notably, \changes{TS, RGI, UCB and KLU} are the only ones of the adaptive designs which
appear unimodal in both scenarios. The unimodality of \changes{TS, RGI, UCB and KLU} under $H_{1}$
(Figure \ref{fig:2arm-hist-rgi-h1}) indicates that, in almost all
realisations of \changes{these trials} under $H_{1}$, the trial is correctly
favouring the superior experimental arm by the end of the trial.

\subsubsection{\label{sub:two-arm-results}Results and discussion}

We now present results of $10^{4}$ repetitions of each trial design using the estimated values described before as \emph{a priori}
critical values.
The results of the simulation are displayed in Table \ref{tab:2arm}. $\mathbb{E}p^{*}$ under
$H_{0}$ is computed as the proportion of patients receiving the control treatment,
and under $H_{1}$ as the proportion of patients receiving the
experimental treatment. The \emph{(s.d.)} values are the standard deviations associated with
each measurement. The \emph{Upper Bound (UB)} row displays a theoretical
optimum for each measurement based on a design which assigns every
patient to the best treatment (\emph{i.e.} $p^{*}=1$) in every trial.

\begin{table}
\noindent\resizebox{\textwidth}{!}{
\begin{tabular}{lc|c||rrlrl||rrlrl|}
 & \multicolumn{1}{c}{} &  & \multicolumn{5}{c||}{$H_{0}:\,\mu_{0}=\mu_{1}=0$} & \multicolumn{5}{c|}{$H_{1}:\,\mu_{0}=0,\,\mu_{1}=0.545$}\tabularnewline
\cline{4-13}
 &  & $C_{\alpha}$ & $\alpha$ & $\mathbb{E}p^{*}$ & (s.d.) & $\mathbb{E}$O & (s.d.) & $(1-\beta)$ & $\mathbb{E}p^{*}$ & (s.d.) & $\mathbb{E}$O & (s.d.)\tabularnewline
\hline
\hline
 & FR & 1.645 & 0.0510 & 0.4997 & (0.05) & -0.0001 & (0.09) & 0.8996 & 0.4997 & (0.05) & 0.2718 & (0.10)\tabularnewline
\hline
(Adaptive & TS & 1.701 & 0.0528 & 0.5006 & (0.11) & 0.0003 & (0.09) & 0.8723 & 0.7317 & (0.10) & 0.3997 & (0.11)\tabularnewline
random) & \changes{TSB}& \changes{1.676} & \changes{0.0519} & \changes{0.4994} & \changes{(0.10)} & \changes{-0.0001} & \changes{(0.09)} & \changes{0.8824} & \changes{0.6962} & \changes{(0.09)} & \changes{0.3816} & \changes{(0.11)}\tabularnewline
\hline
(Semi-random & RBI & 1.998 & 0.0509 & 0.5041 & (0.37) & -0.0001 & (0.09) & 0.3493 & 0.8891 & (0.17) & 0.4845 & (0.13)\tabularnewline
index-based) & RGI & 1.941 & 0.0487 & 0.5005 & (0.27) & 0.0000 & (0.09) & 0.5494 & 0.8764 & (0.09) & 0.4765 & (0.10)\tabularnewline
\hline
\multirow{4}{*}{(Index-based)} & UCB & 2.068 & 0.0508 & 0.5050 & (0.24) & 0.0012 & (0.09) & 0.5575 & 0.8697 & (0.10) & 0.4734 & (0.11)\tabularnewline
 & KLU & 1.867 & 0.0481 & 0.5021 & (0.17) & -0.0001 & (0.09) & 0.7777 & 0.8225 & (0.08) & 0.4489 & (0.10)\tabularnewline
 & CB & 1.782 & 0.0420 & 0.4918 & (0.48) & 0.0007 & (0.09) & 0.1724 & 0.7624 & (0.40) & 0.4139 & (0.24)\tabularnewline
 & GI & 1.951 & 0.0437 & 0.5006 & (0.38) & -0.0010 & (0.09) & 0.2373 & 0.8786 & (0.23) & 0.4796 & (0.16)\tabularnewline
\hline
 & UB &  &  &  &  & 0.0000 & (0.09) &  & 1.0000 & (0.00) & 0.5450 & (0.09)\tabularnewline
\end{tabular}}

\caption{\label{tab:2arm}Comparison \changes{in $10^4$ trial replicates} of operating characteristics of different
two-arm trial designs of size $T=116$, under both hypotheses. $C_{\alpha}:$
critical value used in hypothesis testing; $\alpha$: type I error
rate; $\mathbb{E}p^{*}$: mean proportion of trial patients assigned
the best treatment; (s.d.): standard deviation for each measurement;
$\mathbb{E}$O: mean patient outcome; $(1-\beta)$: statistical power.
UB: theoretical upper bound from assigning all patients best treatment.}
\end{table}

All the adaptive rules achieve better patient welfare than the FR
design \changes{under
$H_{1}$}. \changes{In this scenario RBI, RGI, UCB and GI all perform similarly
well in patient welfare, with $\mathbb{E}$O values between 0.47 and
0.48, the closest values to the theoretical upper bound of 0.545.}
Note that this contrasts with the findings in \cite{villar-a} for the Bernoulli
case, where GI was found to achieve much better patient welfare than
either of the semi-randomised designs. However, \changes{the results for these rules are} in line with their poorer performance in terms of power when compared to the findings in \cite{villar-a}.
\changes{The TS trial is outperformed by the other adaptive designs in terms of patient welfare; this is explained by the tuning parameter $c$ in the TS mechanism which stabilises the randomisation probabilities.} 

The high standard deviations in $p^{*}$ for all the adaptive designs
under $H_{0}$ indicate that $p^{*}$ has a broad distribution across
the realisations of the trial, so the trials are not consistent and
are frequently unbalanced. The standard deviation of $0.48$ for CB
is close to the limiting case where, in each trial, $p^{*}\sim\mathrm{Bernoulli}(\tfrac{1}{2})$,
\emph{i.e.} all patients within a trial are assigned to the same treatment,
which would give a standard deviation of $\sqrt{\tfrac{1}{2}(1-\tfrac{1}{2})}=0.50$
in $p^{*}$ across the trials. This indicates that most trials under
CB (and to some extent also GI and RBI) are highly unbalanced, with
one arm being dropped early on and most patients receiving the same
treatment).

Under $H_{1}$, the high standard deviation in $p^{*}$ under GI arises
from the bimodality observed in Figure \ref{fig:two-arm-hist-gi-h1}:
in a small proportion of realisations of the trial, the control arm
is incorrectly favoured and $p^{*}<\nicefrac{1}{2}$. The lower standard
deviation in $p^{*}$ for RGI confirms that RGI trials are more consistent
in correctly favouring the superior treatment arm. 
As expected, the GI trial has greatly reduced
statistical power (just 24\%) compared to the value of 90\% achieved
by the FR trial. Reduced power is also evident in the other trial
designs (\emph{c.f.} Figure \ref{fig:app-2arm-hist}); CB has
the lowest power (17\%).

\changes{Note that UCB outperforms KLU in patient welfare, but KLU offers significantly higher power (78\%) than UCB (56\%). Interestingly, despite the improved regret bounds for KLU proved in \cite{cappe-et-al} KLU only begins to dominate UCB under both power and patient benefit when the number of patients is very large.  Nevertheless, KLU seems to achieve the best compromise between patient welfare and statistical inference out of other modifications to the UCB algorithm designed to improve regret bounds reviewed for this paper, with power of 78\% and $\mathbb{E}$O of 0.45 (only slightly below the 0.48 achieved by GI). The low standard deviation of 0.10 in expected outcome indicates that the welfare benefit is more consistent than in the GI trial.}

\changes{The results for the batched TS (TSB) illustrate the effects of a blocked implementation of the algorithm to deal with a moderate delay: a marginal increase in power and a considerable decrease of the patient welfare benefits. However, the patient benefit advantages of TSB over FR are considerably large even if assuming a moderate delay in patient recruitment. }

\subsubsection{\label{sub:two-arm-results}Bias in treatment effect estimates}
\begin{figure}[!tbh]
\subfloat[\label{fig:2arm-bias-h1k1}$H_{0}$, control arm $k=0$]{\includegraphics{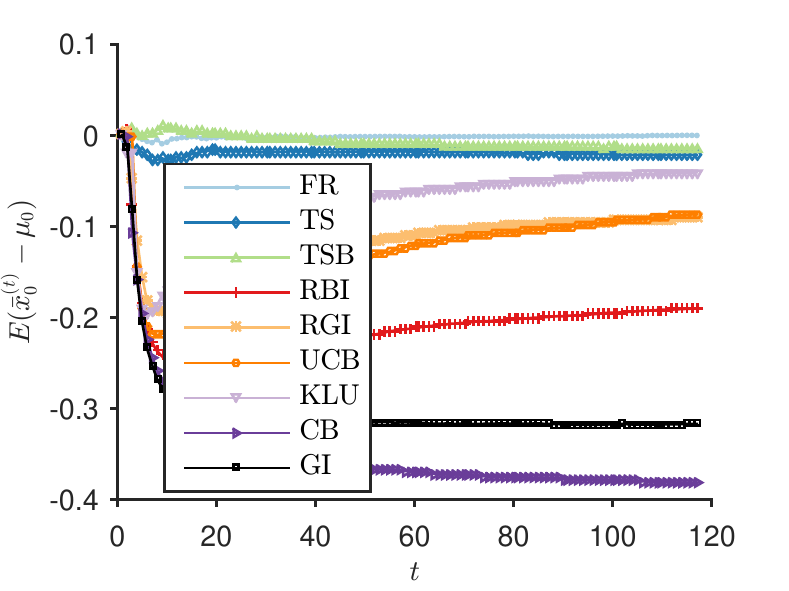}}\hfill{}\subfloat[\label{fig:2arm-bias-h1k2}$H_{0}$, experimental arm $k=1$]{\includegraphics{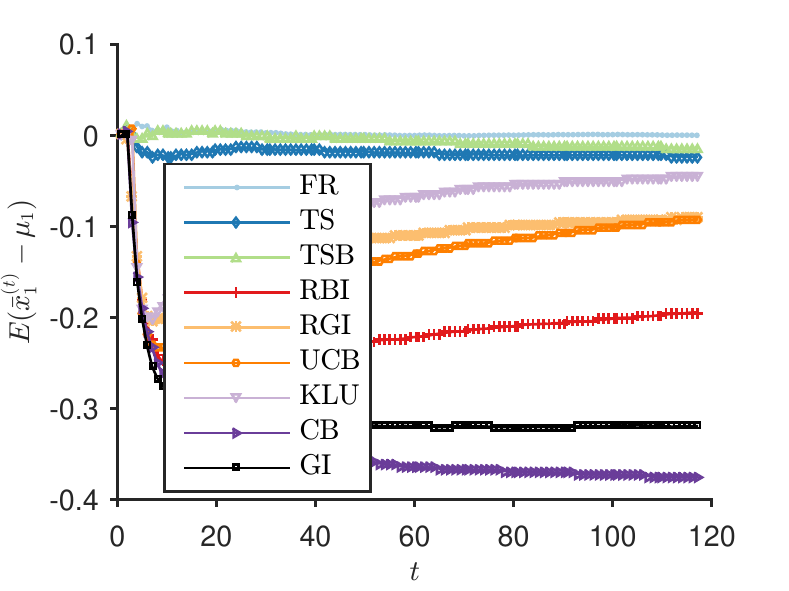}}

\subfloat[\label{fig:2arm-bias-h2k1}$H_{1}$, control arm $k=0$]{\includegraphics{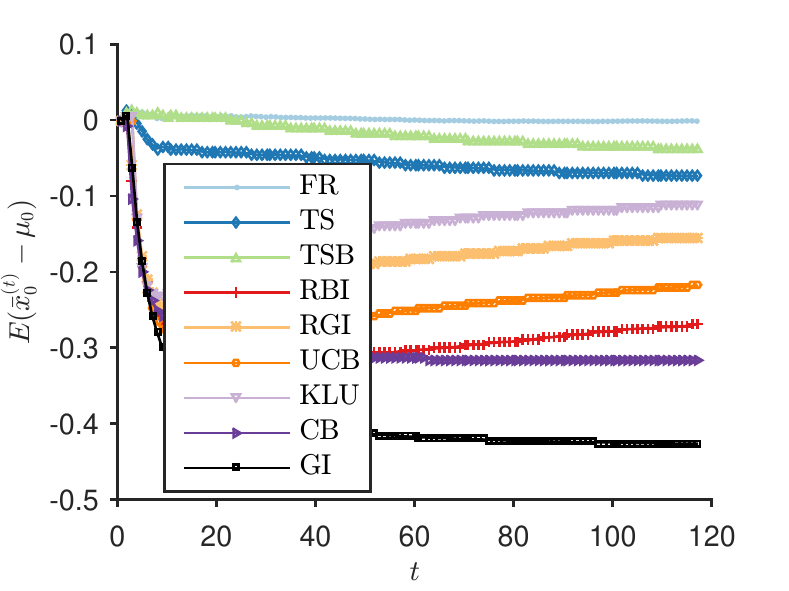}}\hfill{}\subfloat[\label{fig:2arm-bias-h2k2}$H_{1}$, experimental arm $k=1$]{\includegraphics{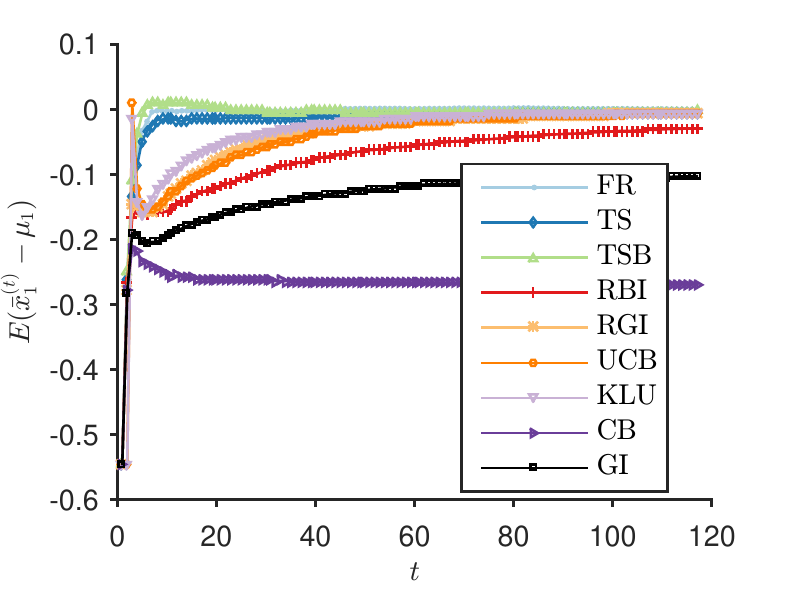}

}

\caption{\label{fig:2arm-bias} $\mathbb{E}(\overline{x}_{k}^{(t)}-\mu_{k})$,
the mean (across the trial realisations) of the bias in the estimated
outcome of each treatment after a total of $t$ patients have been
treated across both arms in the trial, under each scenario (two-arm
trial simulations).}

\end{figure}

Figure \ref{fig:2arm-bias} shows the mean (across the trial realisations)
of the bias $(\overline{x}_{k}^{(t)}-\mu_{k})$ in the estimated outcome
of each treatment after a total of $t$ patients have been treated
across both arms in the trial, under each scenario. Figures \ref{fig:2arm-bias-h1k1}
and \ref{fig:2arm-bias-h1k2} show the GI design introducing a negative
bias into estimates of both treatment's effects; within each trial
realisation this bias will be restricted to one of the two arms, corresponding
to the two modes of the test statistic distribution in Figure \ref{fig:two-arm-hist-gi-h0}.
In all scenarios, the deterministic designs GI and CB exhibit larger
bias than the semi-randomised designs RBI and RGI.


\subsection{\label{sub:4arm-large-results}Four-arm trial scenario}

This scenario uses the TAILoR trial but now considers $K=3$ experimental
treatments to be compared with a control treatment. To achieve a type
I error rate of $5\%$, the critical value is ${C_{\alpha}} =2.0621$ for the FR trials. Once again we
take $\sigma^{2}=1$, and we assume a trial size is of $T=302$ patients since this is the total required
trial size for FR to achieve $(1-\beta)=90\%$ power to detect a difference
of $\delta^{(1)}=0.545$ in treatment outcome.
The trial is implemented under $H_{0}$ with $\mu_{0}=\mu_{1}=\mu_{2}=\mu_{3}=0$
and under $H_{1}$ with $\mu_{0}=0,\,\mu_{1}=\mu_{2}=0.178,\,\mu_{3}=0.545$.
These values are chosen to give the \emph{Least
Favourable Configuration (LFC)} for the trial, with $\mu_{1}=\mu_{2}=\delta^{(0)}$
and $\mu_{3}=\delta^{(1)}$, where, as \cite{wason} explain: ``\emph{\(\delta^{(1)}\) is a prespecified clinically relevant effect, and \(\delta^{(0)}\) is some threshold below which a treatment is considered uninteresting. The configuration is called least favourable as it minimises the probability of recommending a treatment with effect greater than or equal to \(\delta^{(1)}\) amongst all configurations where at least one treatment has a treatment effect of \(\delta^{(1)}\) or higher and no treatment effects lie in the interval \((\delta^{(0)},\delta^{(1)})\).}''
Following \cite{wason}, $\delta^{(0)}=0.178$ is chosen so that the probability
of a patient on a treatment with this treatment effect achieving a
better outcome than a patient on the control treatment is 0.55 and the corresponding probability for
$\delta^{(1)}=0.545$ is 0.65.

We will compare all the trial designs, including now the \emph{Controlled Gittins (CG)} design, in which each
patient is allocated the control treatment with probability $\tfrac{1}{K+1}=0.25$,
and otherwise allocated the drug with the highest value
of the Gittins Index. \changes{We compare CG design against similar procedures: the \emph{Trippa Procedure (TP)} and \emph{Controlled UCB (CUC)} designs. We also include the \emph{Batched Trippa Procedure (TPB)} to assess the effects of delays in outcome observability.}
The Gittins Indices used are 
again based
on discount factor $d=0.995$.
To calculate critical values for the trial designs other than FR,
Monte Carlo simulations were run as explained in section \ref{sub:2arm-MC}. Critical
values are found by calculating the empirical $95^{\mathrm{th}}$-percentile
of the distribution of $Z_{\mathrm{max}}:=\underset{j=1,2,3}{\max}Z_{j}$,
in order to control the FWER. 
Trial simulations are then run
using the computed quantiles as critical values; for each design the
trial is run $10^{4}$ times. 
Results are
displayed in Table \ref{tab:4arm-large}.

\begin{table}[!tbh]
\noindent\resizebox{\textwidth}{!}{
\begin{tabular}{c|c||rrlrl||rrlrl|}
\multicolumn{1}{c}{} &  & \multicolumn{5}{c||}{$H_{0}:\,\mu_{0}=\mu_{1}=\mu_{2}=\mu_{4}=0$} & \multicolumn{5}{c|}{$H_{1}:\,\mu_{0}=0,\,\mu_{1}=\mu_{2}=0.178,\,\mu_{3}=0.545$}\tabularnewline
\cline{3-12}
 & $C_{\alpha}$ & $\alpha$ & $\mathbb{E}p^{*}$ & (s.d.) & $\mathbb{E}$O & (s.d.) & $(1-\beta)$ & $\mathbb{E}p^{*}$ & (s.d.) & $\mathbb{E}$O & (s.d.)\tabularnewline
\hline
\hline
FR & 2.062 & 0.0534 & 0.2493 & (0.02) & 0.0006 & (0.06) & 0.8982 & 0.2502 & (0.02) & 0.2252 & (0.06)\tabularnewline
\hline
TS & 2.198 & 0.0476 & 0.2504 & (0.08) & -0.0001 & (0.06) & 0.8751 & 0.4997 & (0.10) & 0.3394 & (0.07)\tabularnewline
\changes{TSB} & \changes{2.103} & \changes{0.0514} & \changes{0.2497} & \changes{(0.07)} & \changes{-0.0001} & \changes{(0.06)} & \changes{0.8989} & \changes{0.4794} & \changes{(0.10)} & \changes{0.3314} & \changes{(0.07)}\tabularnewline
\hline
RBI & 2.041 & 0.0519 & 0.2469 & (0.27) & 0.0002 & (0.06) & 0.3608 & 0.7917 & (0.22) & 0.4624 & (0.10)\tabularnewline
RGI & 2.070 & 0.0499 & 0.2479 & (0.18) & -0.0006 & (0.06) & 0.6309 & 0.7603 & (0.12) & 0.4462 & (0.07)\tabularnewline
\hline
UCB & 2.223 & 0.0500 & 0.2507 & (0.13) & 0.0001 & (0.06) & 0.7333 & 0.7028 & (0.12) & 0.4238 & (0.08)\tabularnewline
\changes{KLU} & \changes{2.154} & \changes{0.0434} & \changes{0.2502} & \changes{(0.09)} & \changes{-0.0002} & \changes{(0.06)} & \changes{0.8718} & \changes{0.6068} & \changes{(0.10)} & \changes{0.3848} & \changes{(0.07)}\tabularnewline
CB & 1.691 & 0.0524 & 0.2468 & (0.41) & 0.0008 & (0.06) & 0.1075 & 0.4941 & (0.49) & 0.3438 & (0.20)\tabularnewline
GI & 1.955 & 0.0486 & 0.2457 & (0.28) & -0.0008 & (0.06) & 0.2264 & 0.7743 & (0.29) & 0.4552 & (0.13)\tabularnewline
\hline
CG & 1.923 & 0.0405 & 0.4577 & (0.21) & -0.0006 & (0.06) & 0.8667 & 0.5681 & (0.22) & 0.3392 & (0.10)\tabularnewline
\changes{CUC} & \changes{1.934} & \changes{0.0572} & \changes{0.3362} & \changes{(0.09)} & \changes{0.0007} & \changes{(0.06)} & \changes{0.9599} & \changes{0.5357} & \changes{(0.10)} & \changes{0.3277} & \changes{(0.07)}\tabularnewline
TP & 2.027 & 0.0498 & 0.2593 & (0.02) & -0.0010 & (0.06) & 0.9418 & 0.3095 & (0.06) & 0.2462 & (0.06)\tabularnewline
\changes{TPB} & \changes{2.027} & \changes{0.0479} & \changes{0.2488} & \changes{(0.02)} & \changes{0.0003} & \changes{(0.06)} & \changes{0.9342} & \changes{0.3082} & \changes{(0.06)} & \changes{0.2476} & \changes{(0.06)}\tabularnewline

\hline
UB &  &  &  &  & 0.0000 & (0.06) &  & 1.0000 & (0.00) & 0.5450 & (0.06)\tabularnewline
\end{tabular}}

\caption{\label{tab:4arm-large}Comparison \changes{in $10^4$ trial replicates} of operating characteristics of
different four-arm trial designs of size $T=302$, under both hypotheses.
$C_{\alpha}:$ critical value used in hypothesis testing; $\alpha$:
type I error rate; $\mathbb{E}p^{*}$: mean proportion of patients
in a trial assigned the best treatment; (s.d.): standard deviation
for each measurement; $\mathbb{E}$O: mean patient outcome; $(1-\beta)$:
statistical power. UB: theoretical upper bound from assigning all
patients best treatment.}
\end{table}

\changes{As in the two-arm scenario, all the adaptive rules
outperform the FR design under $H_{1}$ in terms of patient welfare, although TP only improves marginally over FR in this case}. The greatest
$\mathbb{E}$O values are achieved by RGI, RBI, UCB and GI, but
these designs and CB exhibit a greatly reduced power level compared with FR,
rendering them less useful as trial designs from a frequentist point
of view. \changes{In particular, CB, which is essentially the simplest myopic approach, exhibits the worst performance in terms of power and variability. As in the two arm trial, KLU achieves considerably greater power than UCB and the welfare benefit is only slightly reduced, offering a very good compromise between the two conflictive objectives.}

\changes{As in the two-armed case TSB (90\%) achieves marginally higher power than TS (88\%) in return for slightly lower patient welfare ($\mathbb{E}$O of 0.33 compared to 0.34). Conversely, TPB results in a slightly reduced power than TP while the patient wlefare is practically identical. In both cases, the difference caused by the moderate ``batching'' of patients' outcomes is small, indicating that 
these adaptive designs could offer patient benefit advantages even if applied without a fully sequential design.}

\changes{As expected, the family of ``protected control'' designs: TP, CG and CUC, offer a compromise between learning (power) and earning (patient welfare).}
Whilst CG does not perform as well as GI in patient welfare, its $\mathbb{E}$O
value of $0.3392$ is still a significant improvement on FR's $0.2252$,
and the $87\%$ power attained by CG is \changes{greater than that of
many of the other adaptive designs}, and only marginally lower than FR's $90\%$. \changes{CUC compares similarly to UCB and dominates over TP by offering a significantly increased patient welfare with a slight increase in power over TP.}
Just as found in \cite{villar-a} \changes{for} the Bernoulli case, fixing the control
allocation in this way is a simple heuristic modification of adaptive allocation rules that results in good trial designs in terms of
both patient welfare and frequentist operating characteristics.

\begin{figure}[!tbh]
\subfloat[\label{fig:4armlarge-bias-h1k1}$H_{0}$, control arm $k=0$]{\includegraphics{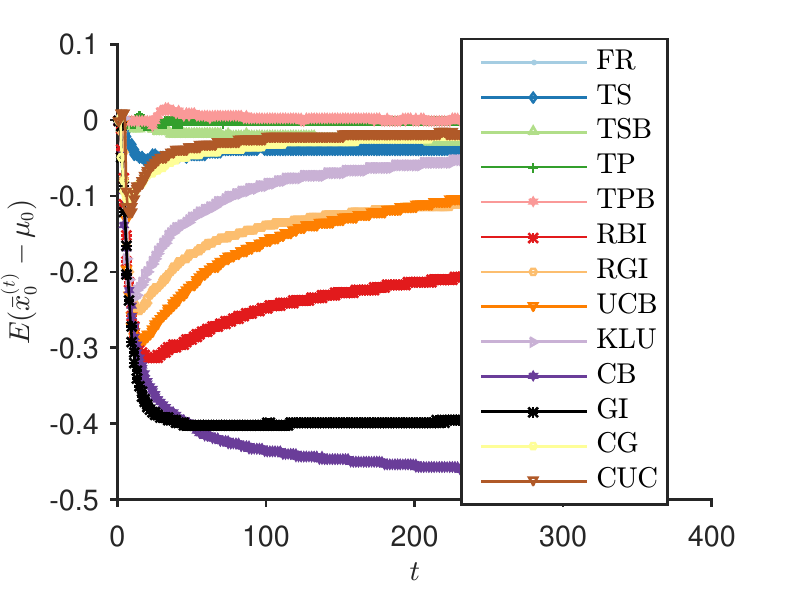}}\hfill{}\subfloat[\label{fig:4armlarge-bias-h1k4}$H_{0}$, experimental arm $k=3$]{\includegraphics{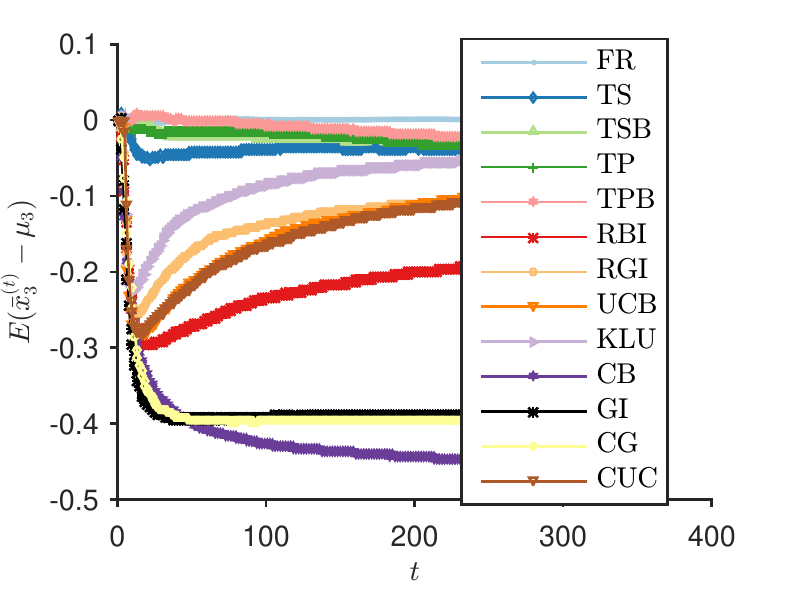}}

\subfloat[\label{fig:4armlarge-bias-h2k1}$H_{1}$, control arm $k=0$]{\includegraphics{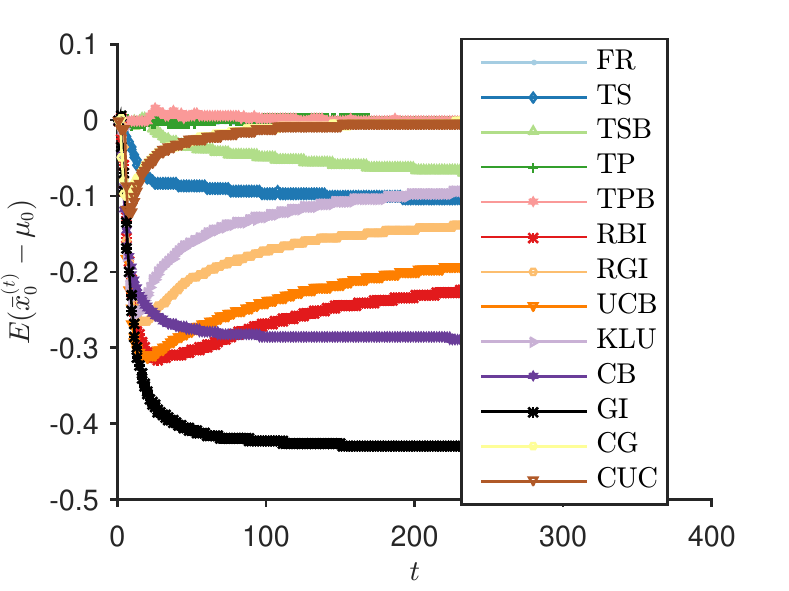}}\hfill{}\subfloat[\label{fig:4armlarge-bias-h2k4}$H_{1}$, experimental arm $k=3$]{\includegraphics{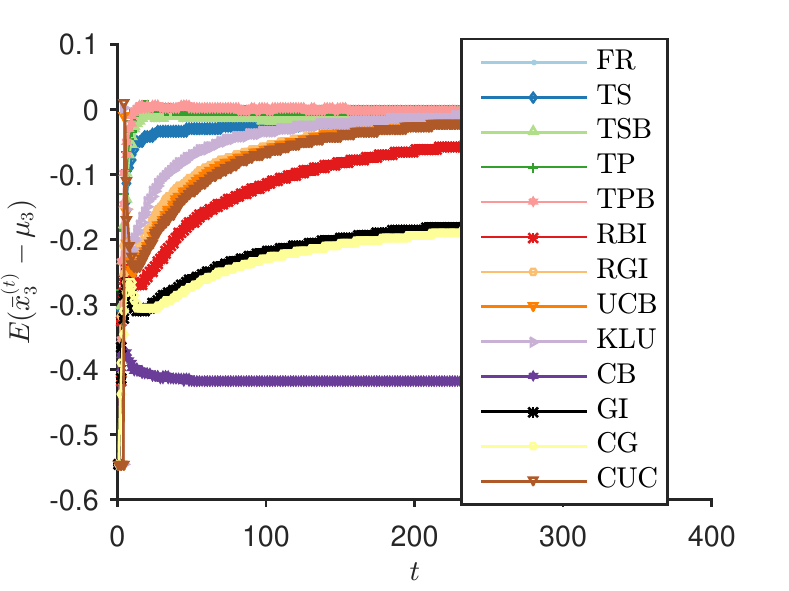}}

\caption{\label{fig:4armlarge-bias} $\mathbb{E}(\overline{x}_{k}^{(t)}-\mu_{k})$,
the mean (across the trial repeats) of the bias in the estimated treatment
outcome of each drug under each scenario in the four arm trial (large
sample size)}
\end{figure}

Figure \ref{fig:4armlarge-bias} shows the bias in the estimates of
treatment outcomes in this scenario, for the control treatment and
the best experimental treatment (arm $3$). For the designs which
were included in the two-arm simulation, the results here are similar.
CG significantly lowers the bias in the estimates of control treatment
outcomes, but it does not improve the issue of negatively biased estimates
of unselected experimental treatment outcomes, where it performs almost
identically to the original GI.

\subsection{\label{sub:4arm-rare-results}Four-arm rare disease trial scenario}

The final simulation scenario is the same as in section \ref{sub:4arm-large-results}
but with the trial size reduced to $T=64$ to imitate a rare disease
setting where the number of patients who can be recruited is limited. Notice that for $T=64$ the FR trial will achieve a power of 30\%
while controlling the FWER within 5\%.
The same critical values are used for hypothesis testing as in the
large trials in section \ref{sub:4arm-large-results}, based on the assumption
that (especially in a trial where patients are recruited sequentially)
the experimenter might not know at the start of the trial the total
number of patients she will be able to recruit, so more appropriate
critical values cannot be estimated \emph{a priori}.
Based on the same reasoning  we continue to use the original
choice of $d=0.995$. %
%

\begin{table}[!tbh]
\noindent\resizebox{\textwidth}{!}{
\begin{tabular}{c|c||rrlrl||rrlrl|}
\multicolumn{1}{c}{} & \multicolumn{1}{c||}{} & \multicolumn{5}{c||}{$H_{0}:\,\mu_{0}=\mu_{1}=\mu_{2}=\mu_{4}=0$} & \multicolumn{5}{c|}{$H_{1}:\,\mu_{0}=0,\,\mu_{1}=\mu_{2}=0.178,\,\mu_{3}=0.545$}\tabularnewline
\cline{3-12}
 & $C_{\alpha}$ & $\alpha$ & $\mathbb{E}p^{*}$ & (s.d.) & $\mathbb{E}$O & (s.d.) & $(1-\beta)$ & $\mathbb{E}p^{*}$ & (s.d.) & $\mathbb{E}$O & (s.d.)\tabularnewline
\hline
\hline
FR & 2.062 & 0.0490 & 0.2497 & (0.05) & 0.0009 & (0.13) & 0.2975 & 0.2490 & (0.05) & 0.2260 & (0.13)\tabularnewline
\hline
TS & 2.198 & 0.0510 & 0.2491 & (0.09) & -0.0001 & (0.13) & 0.2592 & 0.3594 & (0.11) & 0.2779 & (0.13)\tabularnewline
\changes{TSB} & \changes{2.103} & \changes{0.0499} & \changes{0.2489} & \changes{(0.08)} & \changes{-0.0010} & \changes{(0.13)} & \changes{0.2901} & \changes{0.3245} & \changes{(0.09)} & \changes{0.2625} & \changes{(0.13)}\tabularnewline
\hline
RBI & 2.041 & 0.0471 & 0.2497 & (0.22) & 0.0000 & (0.13) & 0.1619 & 0.5351 & (0.26) & 0.3529 & (0.16)\tabularnewline
RGI & 2.070 & 0.0565 & 0.2489 & (0.14) & 0.0013 & (0.12) & 0.2344 & 0.4725 & (0.18) & 0.3258 & (0.15)\tabularnewline
\hline
UCB & 2.223 & 0.0444 & 0.2475 & (0.14) & -0.0028 & (0.13) & 0.1730 & 0.4772 & (0.18) & 0.3296 & (0.15)\tabularnewline
\changes{KLU} & \changes{2.154} & \changes{0.0503} & \changes{0.2492} & \changes{(0.10)} & \changes{-0.0023} & \changes{(0.13)} & \changes{0.2452} & \changes{0.4194} & \changes{(0.13)} & \changes{0.3043} & \changes{(0.13)}\tabularnewline
CB & 1.691 & 0.0515 & 0.2522 & (0.37) & -0.0006 & (0.12) & 0.0775 & 0.4569 & (0.46) & 0.3239 & (0.22)\tabularnewline
GI & 1.955 & 0.0477 & 0.2529 & (0.24) & 0.0015 & (0.13) & 0.1226 & 0.5445 & (0.29) & 0.3585 & (0.17)\tabularnewline
\hline
CG & 1.923 & 0.0523 & 0.4100 & (0.17) & -0.0002 & (0.12) & 0.3806 & 0.4061 & (0.22) & 0.2742 & (0.15)\tabularnewline
\changes{CUC} & \changes{1.934} & \changes{0.0573} & \changes{0.3445} & \changes{(0.10)} & \changes{-0.0011} & \changes{(0.13)} & \changes{0.3851} & \changes{0.3821} & \changes{(0.15)} & \changes{0.2670} & \changes{(0.14)}\tabularnewline
TP & 2.027 & 0.0462 & 0.2275 & (0.04) & 0.0011 & (0.12) & 0.3174 & 0.3256 & (0.10) & 0.2534 & (0.13)\tabularnewline
\changes{TPB} & \changes{2.027} & \changes{0.0472} & \changes{0.1751} & \changes{(0.04)} & \changes{-0.0019} & \changes{(0.12)} & \changes{0.2674} & \changes{0.3141} & \changes{(0.07)} & \changes{0.2620} & \changes{(0.13)}\tabularnewline
\hline
UB &  &  &  &  & 0.0000 & (0.13) &  & 1.0000 & (0.00) & 0.5450 & (0.13)\tabularnewline
\end{tabular}}

\caption{\label{tab:4arm-rare}Comparison \changes{in $10^4$ trial replicates} of operating characteristics of different
four-arm trial designs of size $T=64$, under both hypotheses. $\alpha$:
type I error rate; $\mathbb{E}p^{*}$: mean proportion of patients
in a trial assigned the best treatment; (s.d.): standard deviation
for each measurement; $\mathbb{E}$O: mean patient outcome; $(1-\beta)$:
statistical power. UB: theoretical upper bound from assigning all
patients best treatment.}
\end{table}

Table \ref{tab:4arm-rare} shows the full results of the simulations.
Due to the greatly reduced sample sizes, all designs now achieve much
lower power\changes{, a common situation in drug development for rare diseases}. \changes{In a situation where $N\gg T$, statistical power is important,
and CUC and CG offer the best compromise. Both perform similarly well, achieving higher power
than FR, and offering a marked improvement in patient welfare
compared with FR}. However, if the trial subjects comprise most of
the total population to be treated ($\nicefrac{T}{N}\approx1$), then
\changes{GI and RBI provide} the best patient outcome throughout the trial.

\changes{The results in the table for the batched approaches show that, as expected, as the delay in recruitment is more severe the advantages of TSB and TPB over FR are significantly reduced (though both designs still offer important patient welfare advantages). Noticeably, the effect on power and patient welfare of a severe delay in the controlled version (i.e. TPB) differs to that of the \emph{uncontrolled} variant (TSB). The controlled version has its power levels reduced as the delay increases (while the opposite happens to TSB).  TP improves power over FR by matching the allocation of the control arm to that of the best performing arm, therefore increasing the allocation to these two arms over the other arms. With a larger delay TBP will allocate larger number of patients to all arms which therefore reduces its marginal power levels compared to TP. For TSB the power improvement is explained because the design cannot skew allocation to the best arm as fast as with TS, thus allocating more patients to all arms when compared to TS.  }

One distinctive feature of the results is that the Type I error rate
$\alpha$ in the UCB trial is lower than the expected $5\%$,
at just $4.4\%$. As explained above, the same critical values for
hypothesis testing have been used as in §\ref{sub:4arm-large-results},
since the experimenter might not known in advance the total number
of patients to be recruited. Figure \ref{fig:critical-values-trial-size}
shows how the appropriate critical value $C_{0.05}$ for hypothesis
testing with a Type I error rate $\alpha=5\%$ varies according to
the size $T$ of the trial. For most trial designs, there is little
variation in $C_{0.05}$ as $T$ increases. However, for the UCB trial,
$C_{0.05}$ increases significantly with $T$; as a result, the appropriate
critical value to ensure a 5\% Type I error rate is lower for
the smaller 64 person trial, at $C_{0.05}=2.10$, compared to $C_{0.05}=2.22$
for the 302 person trial. Therefore, the 64 person trial conducted
at the higher critical value of $2.22$ generates a low Type
I error rate, and the power is even lower than it could be if the
test was relaxed by lowering the critical value to $2.10$. As a result,
the UCB mechanism may be unsuitable for trials where the total number
of patients to be recruited is not known in advance. \changes{This effect is less pronounced in the KLU variant making it more suitable in that case.}
\begin{figure}[!tbh]
\includegraphics{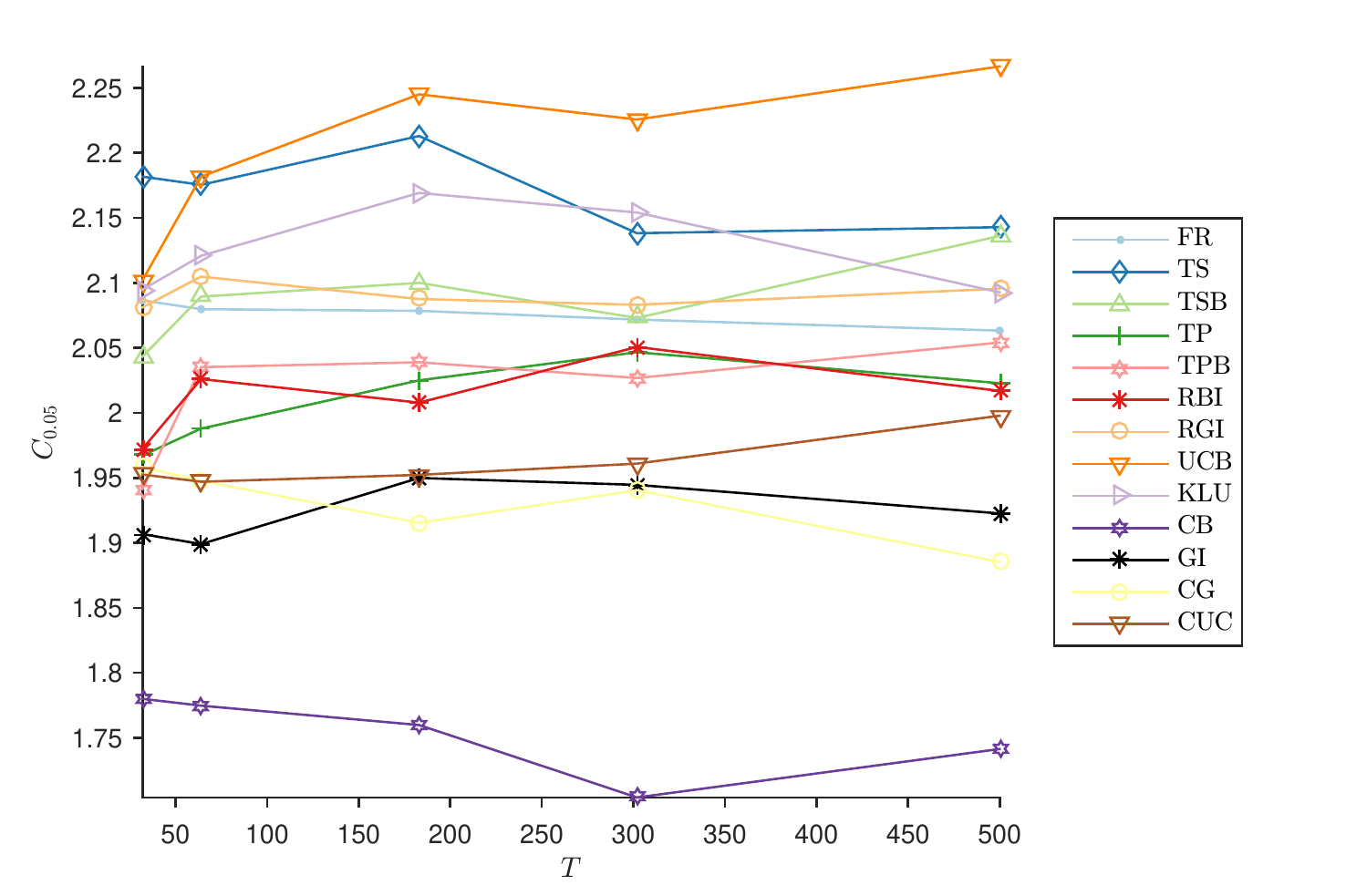}\caption{\label{fig:critical-values-trial-size}Empirical critical values $C_{0.05}$
for one-tailed testing to maintain 5\% FWER in the four arm trial
design, against number $T$ of patients in the trial}

\end{figure}

Since the trial size was much smaller than expected, there is a motivation
to consider if using a smaller value for $d$ would affect results, as a smaller discounting factor
corresponds to putting less value on learning for the future. Note that, when varying the discount factor, we might expect
the distribution of the test statistic $\mathbf{Z}$ to change, and
so critical values for the hypothesis tests would have to be recalculated
for each discount factor for the Gittins index designs, via a Monte
Carlo simulation as in §\ref{sub:4arm-large-results}.  In simulations not included here we found that for this trial setting in all of GI, RGI and CG 
there is no significant variation in patient outcome between discount factors in $\{0.9, 0.95, 0.995, 0.99\}$.

\section{\label{sec:conclusion}Conclusions and discussion}

The simulation results provided by this paper illustrate  how the
index-based response-adaptive design derived from the 
MABP can lead to significant improvements in patient welfare
also 
with a normally distributed endpoint. In
all situations, designs based on the Gittins index achieved \changes{the largest
patient welfare gain over}  FR trials or myopic designs \changes{currently in use in drug development such as TP}. However, there
are a number of limitations to the effectiveness of the purely deterministic
Gittins index design that still prevail. \changes{As in the binary case,} the Gittins index rule exhibits considerably lower power than
FR, and whilst the loss of power can be alleviated to some extent
by the introduction of random perturbations to the indices (RGI),
in the two-arm trial the power achieved is still not sufficient for
most clinical trials unless the exploration term is correctly calibrated.

\changes{In a multi-armed case, the patient welfare advantages of adaptive designs, and GI-based particularly, over FR are the largest. Moreover, there are adaptive designs that can offer more power than FR together with a patient benefit advantage, making them suitable for drug development for common conditions.   }
In the four-arm case based on a real trial we studied, a small deviation from
optimality by \changes{protecting} the allocation of the control treatment (CG \changes{and CUC})
offers a power close to (or even above) FR's while still providing considerable patient
benefit. In contexts where power is relatively less important (if there are very
few disease sufferers outside the trial), GI, RGI or UCB offer even better
patient welfare \changes{at the expense of a power reduction}.

\changes{
There are designs that increase power levels of the UCB algorithm by introducing modifications to improve its asymptotic regret bounds, as shown for KLU in \cite{cappe-et-al}. However, such power gains require a very large number of patients in the trial to be also accompanied by similar patient welfare advantages. For example,  KLU dominates over UCB under both criteria only in scenarios where trials had more than thousands of patients. For smaller (and more realistic) trial sizes, as the ones considered in this paper, UCB had better patient welfare and less power than KLU. Nevertheless, rules like KLU offer a good trade-off between the two objectives and can be suitable designs for common diseases. }

An important observation drawn from the simulations provided by this paper is that
the type I error deflation of the GI observed for the Bernoulli case does not hold in the normally distributed case. Actually, if no correction is introduced using a standard test will result in an important type I error inflation. In this work we have outlined a simulation based procedure that can be used to prevent this inflation.

As pointed out in \cite{berry-1989}, trying to shoehorn trials employing an
adaptive design from a Bayesian viewpoint into traditional frequentist
hypothesis tests may not be the most appropriate method of inference.
Alongside the statistical community's faith in randomisation is a
trust in frequentist inference, so this is generally used even in
Bayesian trials to make the results as persuasive as possible. But,
the inferential power \changes{and the potential patient benefit} from adaptive trials could be improved
by applying Bayesian inference methods \changes{combined with the use of prior data}. Further research could seek an appropriate method of Bayesian inference
based on index-based adaptive trials, \emph{e.g.} by considering which
arm the adaptive design is favouring most at the end of the trial,
or by incorporating information derived from historical data.%

None of the Bayesian allocation mechanisms considered here manages
to completely eliminate the statistical bias phenomenon; further research
is needed to seek an alternative mechanism or a means of accounting
for the bias introduced. Moreover, they all carry a level of selection
bias which, while not studied in the simulations included in this paper, could
lead to much greater bias in clinical trials on a real population.
Further research is needed to investigate whether significant practical
problems will arise from selection bias in real trials, and whether
random perturbations to the indices are sufficient to eliminate these
problems. Alternatively, to overcome this limitation further research could repeat the idea introduced in \cite{villar-b} to randomise group of patients based on probabilities determined by the Gittins Indices for trials with
continuous endpoints. \changes{For the procedures that protect allocation to the control arm we recommend a randomised implementation (where a patient is randomised to control or experimental arms with probabilities $1/(K+1)$, $1-1/(K+1)$ respectively and then allocated to experimental arms according to the index rule). A systematic allocation to control arm (i.e., 1 in every $K+1$ patient is allocated to control) while in theory is equivalent to its randomised counterpart in practice is subject to a very high degree of selection bias. }

Some adaptive trials are designed to take account of covariates
in the trial population (\emph{e.g. }age, weight, blood pressure)
which might affect the treatment response, by ensuring the allocations
are balanced across the covariate factors
{\cite{atkinson-biswas}
}. \changes{Other trials incorporating covariate information combined with response-adaptive procedures with the aim of identifying superior treatments more quickly, mainly treatments that work better within subgroups, is an essential requirement to make \emph{personalised medicine} possible.}
}  Some work has been done on incorporating covariates into the one-armed
bandit problem, \changes{yet further research is needed to extend the approach to multi-armed
bandits used in this work to clinical trials with biomarkers}. See \cite{woodroofe}, \cite{sarkar} and \changes{\cite{perchet-rigollet}}.
}

\changes{None of the adaptive designs considered formally accounts for the estimated population size. The index-based approaches indirectly can consider that by appropriate selection of the discount factor. However, the results in this paper suggest that the choice between implementing a traditional FR design or an adaptive design should depend on the current belief of how large the population of patients outside the trial is. }

Finally, \changes{the results presented in this paper have highlighted that} further analogous research is needed to extend these results \changes{and address potential specific issues} to trials with other endpoints, such as continuous endpoints that are not normally distributed. 

\subsection{Acknowledgements}

This work was funded by the Biometrika Trust and the UK Medical Research Council (grant number MC\_UP\_1302/2).
\clearpage

\section*{\label{sub:fwer}Appendix I: Controlling the family-wise type I error rate}

%
In order to control the FWER when carrying out multiple testing, we
need to consider the joint distribution of $Z_{1},\ldots,Z_{k}$.
We have, for $1\le j\neq k\le K$,
\[
\begin{array}{rcl}
\mathrm{cov}\left(Z_{j},Z_{k}\right) & = & \mathrm{cov}\left(\frac{\overline{X}_{j}-\overline{X}_{0}}{\sigma\sqrt{\tfrac{1}{n_{j}}+\tfrac{1}{n_{0}}}},\frac{\overline{X}_{k}-\overline{X}_{0}}{\sigma\sqrt{\tfrac{1}{n_{k}}+\tfrac{1}{n_{0}}}}\right)\\
\\
 & = & \frac{1}{\sigma^{2}}\frac{1}{\sqrt{\tfrac{1}{n_{j}}+\tfrac{1}{n_{0}}}}\frac{1}{\sqrt{\tfrac{1}{n_{k}}+\tfrac{1}{n_{0}}}}\mathrm{Var}\left(\overline{X}_{0}\right)
\end{array}
\]
by the independence of $Z_{j}$ and $Z_{k}$. Using the fact that
$\mathrm{Var}\left(\overline{X}_{0}\right)=\tfrac{\sigma^{2}}{n_{0}}$,
\[
\mathrm{cov}\left(Z_{j},Z_{k}\right)=\left[\left(1+\tfrac{n_{0}}{n_{j}}\right)\left(1+\tfrac{n_{0}}{n_{k}}\right)\right]^{-\tfrac{1}{2}}.
\]
Error rates are lowest when the variance of the sample means is minimised,
which corresponds to the trial being well balanced: in a RCT trial with fixed equal randomisation
all the sample sizes are (asymptotically) equal, so we will have
a good approximation for $n_{0}\approx n_{1}\approx\cdots\approx n_{K}$
and $\mathrm{cov}\left(Z_{j},Z_{k}\right)\approx\tfrac{1}{2}$. Hence,
under $H_{0,G}$, $\delta_{1}=\cdots=\delta_{K}=0$ and
\[
\mathbf{Z}=(Z_{1},\ldots,Z_{K})\sim N_{K}\left(0,\Sigma_{K}\right)
\]
 where $\Sigma_{K}$ is the $K\times K$ matrix given by
\[
\left(\Sigma_{K}\right)_{ij}=\begin{cases}
1, & i=j\\
\tfrac{1}{2}, & i\neq j
\end{cases}.
\]
So we would expect a RCT trial to control the FWER at level $\alpha$
by using critical value \global\long\def\ca{C_{\alpha}}
$\ca$ satisfying
\begin{equation}
\underset{-\infty}{\overset{C_{\alpha}}{\int}}\cdots\underset{-\infty}{\overset{C_{\alpha}}{\int}}\phi_{\Sigma_{K}}(x_{1},\ldots,x_{K})\mathrm{d}x_{1}\ldots\mathrm{d}x_{K}=1-\alpha\label{eq:normal-fwer}
\end{equation}
where $\phi_{\Sigma_{K}}$ is the probability density function of
a multi-variate normal $N_{K}(0,\Sigma_{K})$ distribution, \emph{i.e.}
ensuring that
\begin{equation}
\mathbb{P}\left[\underset{k=1,\ldots,K}{\max}Z_{k}\le C_{\alpha}\right]=1-\alpha.\label{eq:fwer-zmax}
\end{equation}

\section*{\label{sub:fwer}Appendix II: Calculating empirical cut-off values to control the type I error rate}

\begin{figure}[!htbp]

\vspace{-0.5cm}

\subfloat[TS trial under $H_{0}$]{\includegraphics{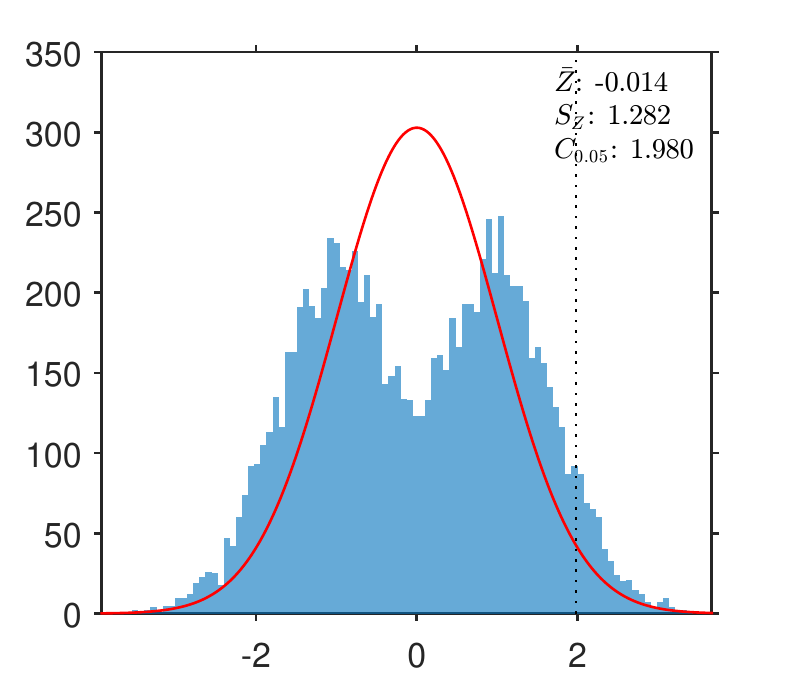}}\hfill{}
\subfloat[\label{fig:2arm-hist-rgi-h1}TS trial under $H_{1}$]{\includegraphics{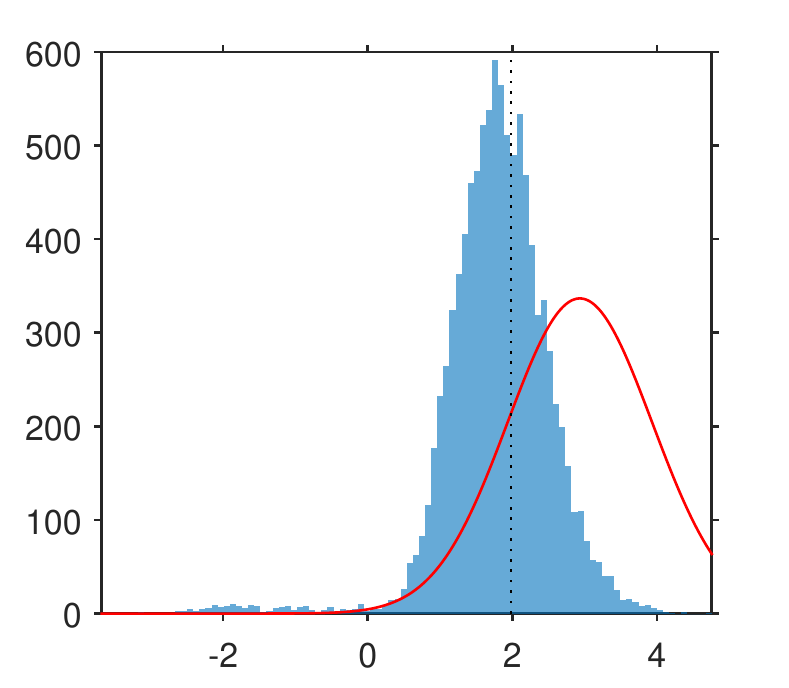}}
\vspace{-0.1cm}
\subfloat[RBI trial under $H_{0}$]{\includegraphics{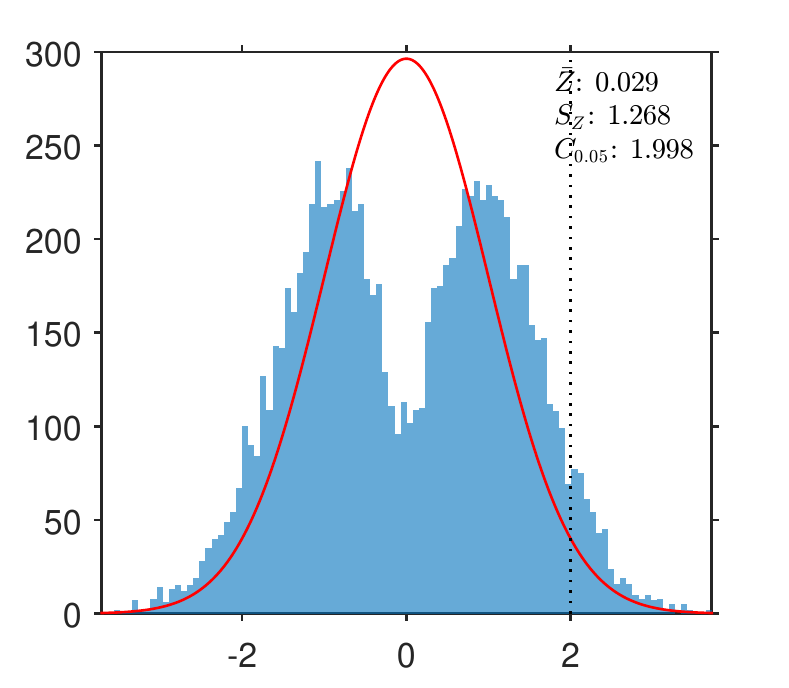}}\hfill{}
\subfloat[RBI trial under $H_{1}$]{\includegraphics{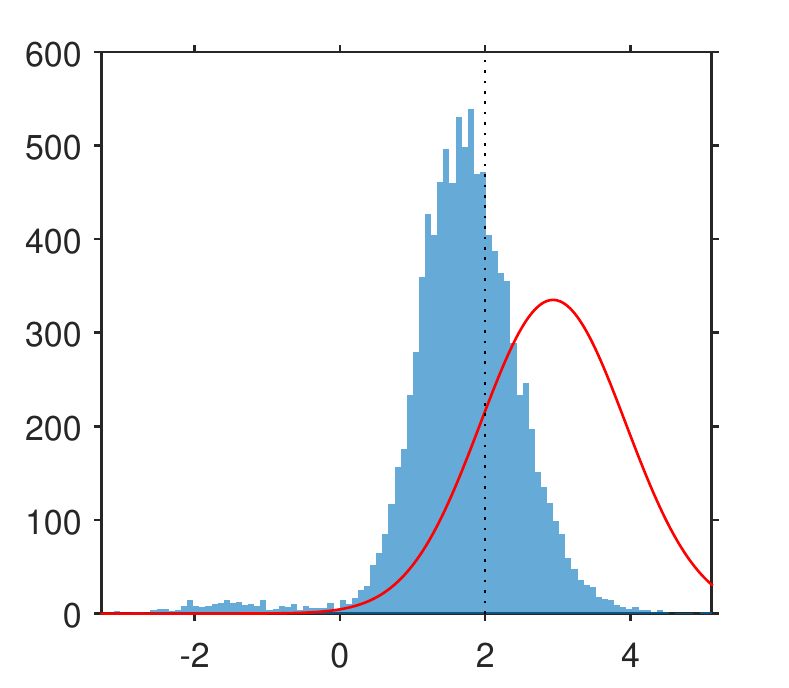}}
\vspace{-0.1cm}
\subfloat[RGI trial under $H_{0}$]{\includegraphics{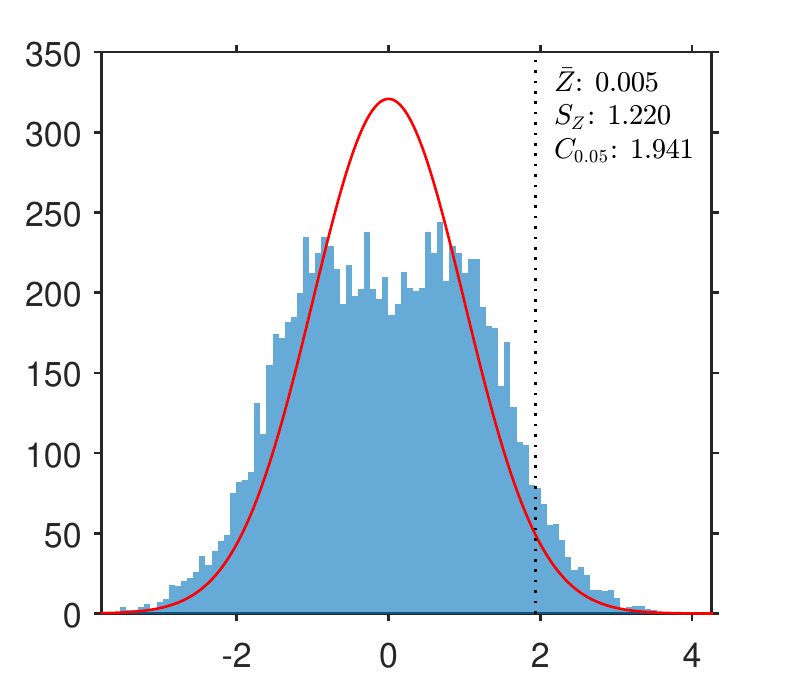}}\hfill{}
\subfloat[RGI trial under $H_{1}$]{\includegraphics{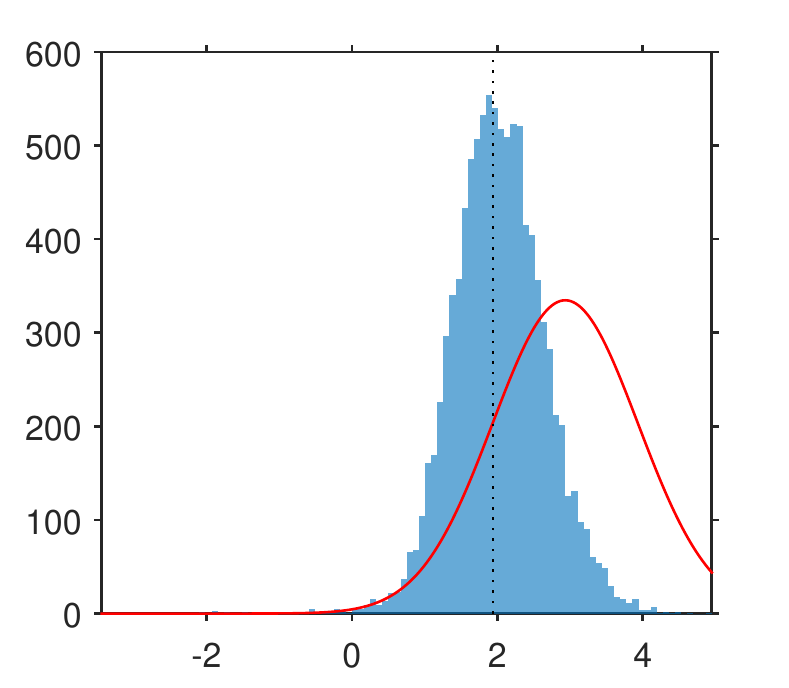}}

\caption{\label{fig:app-2arm-hist}Histograms of empirical distributions of
the test statistic $Z_{0,1}$ in \changes{TS, RBI, RGI, UCB, KLU and CB} two-arm
trials, implemented under each hypothesis (as in Figure \ref{fig:two-arm-hist}).
Also marked is the standard normal distribution which $Z_{0,1}$ should
follow in the FR trial (red). For each design, the sample mean $\bar{Z}_{0,1}$,
standard deviation $S_{Z_{0,1}}$ and an empirical $95^{\mathrm{th}}$-percentile
$C_{0.05}$ have been calculated under $H_{0}$. The empirical $95^{\mathrm{th}}$-percentile
under $H_{0}$ will correspond to the critical value for hypothesis
testing, and is marked by a vertical dotted line on the histograms.}
\end{figure}

\begin{figure}[!htbp]
\ContinuedFloat

\subfloat[UCB trial under $H_{0}$]{\includegraphics{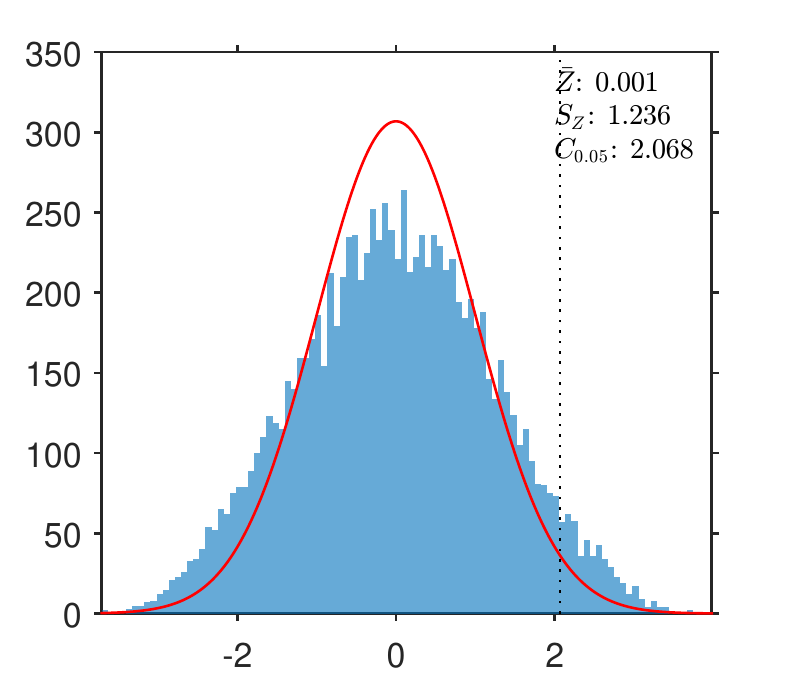}}\hfill{}
\subfloat[UCB trial under $H_{1}$]{\includegraphics{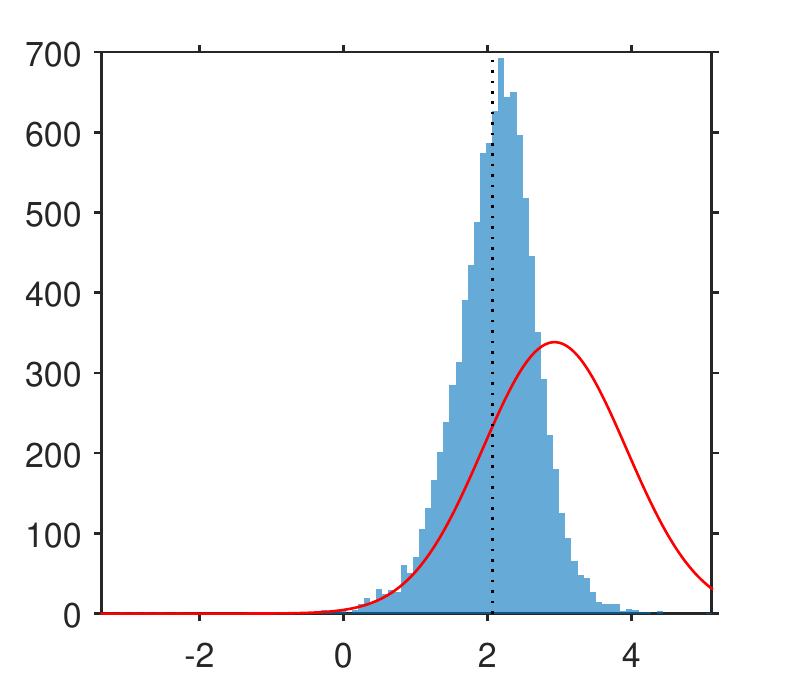}}
\vspace{-0.08cm}
\subfloat[KLU trial under $H_{0}$]{\includegraphics{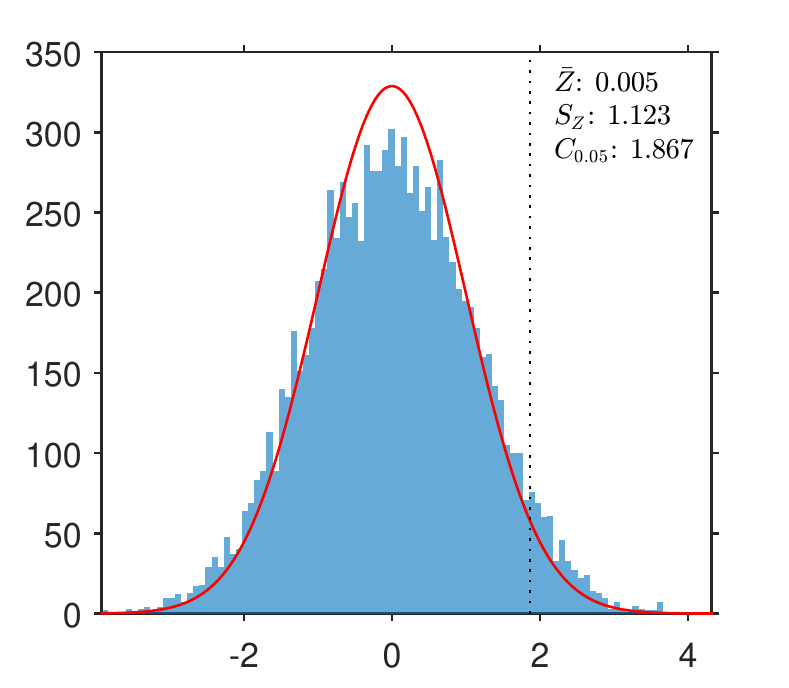}}\hfill{}
\subfloat[KLU trial under $H_{1}$]{\includegraphics{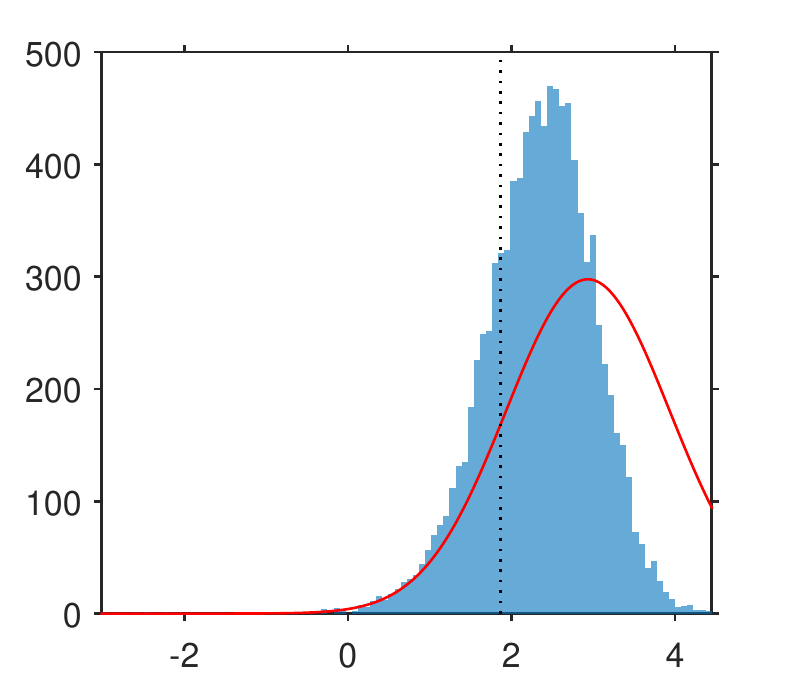}}
\vspace{-0.2cm}
\subfloat[CB trial under $H_{0}$]{\includegraphics{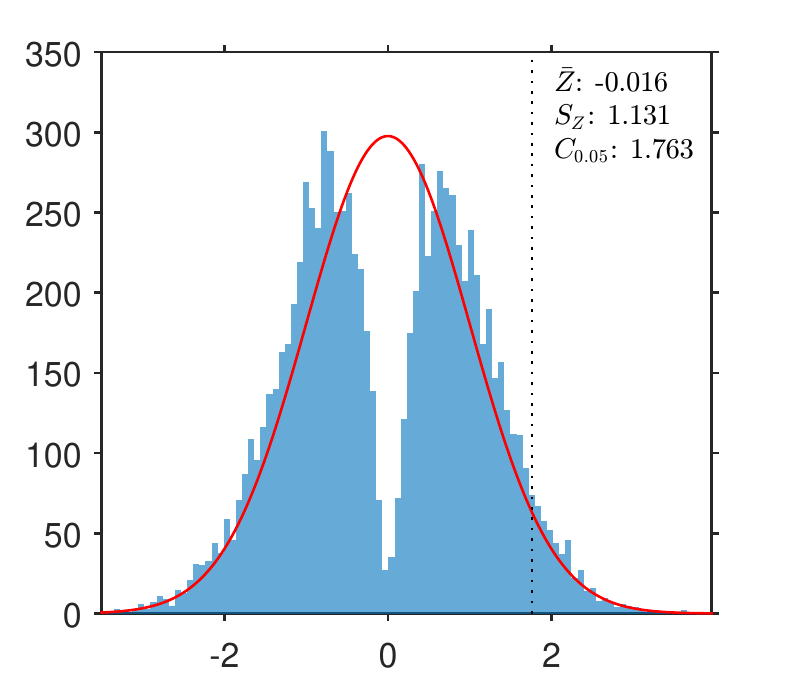}}
\hfill{}\subfloat[CB trial under $H_{1}$]{\includegraphics{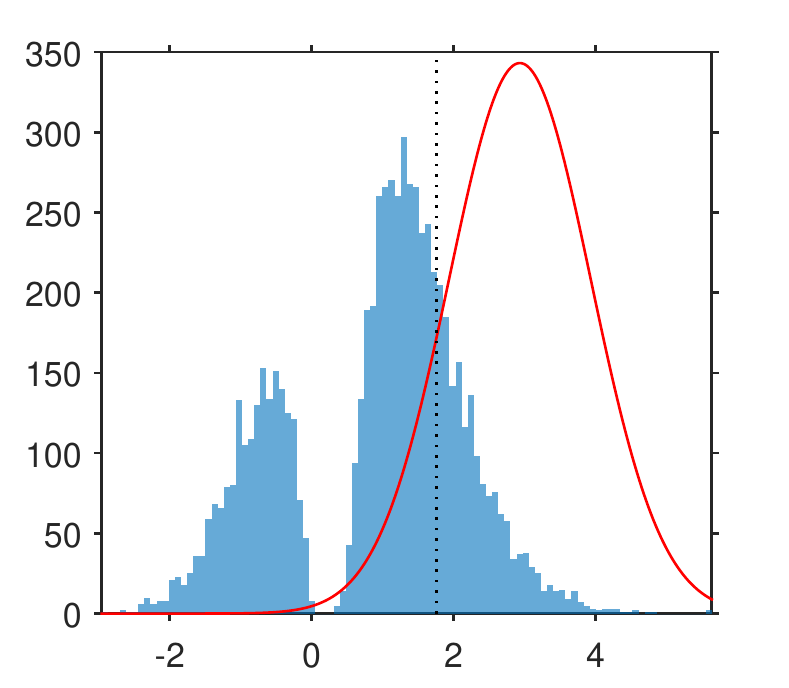}}

\phantomcaption
\end{figure}

\end{document}